\documentclass[journal,transmag]{IEEEtran}
\ifCLASSINFOpdf
\else
\fi
\usepackage{graphicx}

\usepackage{amsmath}
\usepackage{amssymb}
\usepackage{amsthm}
\usepackage{enumerate}
\usepackage{mathtools}
\usepackage{xcolor}
\usepackage{xfrac}

\usepackage{float}
\usepackage{subfig}

\newtheorem{theorem}{Theorem}

\newtheorem{definition}{Definition}

\newtheorem{prpty}{Property}
\newtheorem{example}{Example}

\usepackage{float}
\usepackage{subfig}

\usepackage{algorithm}
\usepackage{algpseudocode}

\makeatletter
\let\OldStatex\Statex
\renewcommand{\Statex}[1][3]{%
  \setlength\@tempdima{\algorithmicindent}%
  \OldStatex\hskip\dimexpr#1\@tempdima\relax}
\makeatother

\makeatletter
\newcommand{\StatexIndent}[1][3]{%
  \setlength\@tempdima{\algorithmicindent}%
  \Statex\hskip\dimexpr#1\@tempdima\relax}
\algdef{S}[IF]{IfNoThen}[1]{\algorithmicif\ #1}%
\makeatother

\newcommand{\p}{\partial}
\newcommand{\myth}{^\textup{th}}

\newcommand{\real}{\mathbb{R}}
\newcommand{\sym}{\mathbb{S}}

\newcommand{\tx}{\tilde{x}}
\newcommand{\tu}{\tilde{u}}
\newcommand{\ty}{\tilde{y}}

\newcommand{\lse}{J^0}
\newcommand{\se}{J}

\newcommand{\lrlse}{\hat{J}_\lambda}
\newcommand{\llse}{{J}_\lambda}

\newcommand{\modvar}{\rho}

\newcommand{\fullvar}{\theta}
\newcommand{\decvar}{\nu}

\newcommand{\liftvar}{\modvar}

\newcommand{\postol}{\mu}

\newcommand{\stblset}{\mathcal{P}}
\newcommand{\sosset}{\Theta}

\newcommand{\sosmat}{S}
\newcommand{\ltistblset}{\sosset_l}

\newcommand{\psos}{p}
\newcommand{\asos}{A_s}

\newcommand{\monos}{\omega}

\newcommand{\ddM}{{W}}
\newcommand{\ddm}{w}

\newcommand{\ours}{our proposed algorithm} 
\newcommand{\mysection}{{Section }} 
\newcommand{\refsec}{\S} 

\newcommand{\cfont}[1]{{\small\texttt{#1}}}

\newcommand{\mtovec}[1]{{\textup{vec}(#1)}}

\newcommand{\sig}{\emph{Sig}}
\newcommand{\sigss}{\emph{Sig}$^*$} 
\newcommand{\wave}{\emph{Wav}}
\newcommand{\wavess}{\emph{Wav$^*$}}

\newcommand{\tte}{\times 10}

\newcommand{\ml}{\prec}
\newcommand{\mleq}{\preceq}
\newcommand{\mg}{\succ}
\newcommand{\mgeq}{\succeq}
\newcommand{\mng}{\nsucc}

\newcommand{\defeq}{:=}

\newcommand{\xapa}{\mathcal{A}}
\newcommand{\xapb}{\mathcal{B}}

\newcommand{\regp}{\delta}

\begin{document}
%
\title{Specialized Interior Point Algorithm for \\ Stable Nonlinear System Identification}


\author{\IEEEauthorblockN{Jack Umenberger\IEEEauthorrefmark{1} and
Ian R. Manchester\IEEEauthorrefmark{1}}

\IEEEauthorblockA{\IEEEauthorrefmark{1}Australian Centre for Field Robotics,
The University of Sydney, NSW 2006 Australia}

\thanks{Manuscript received April 14, 2017. 
Corresponding author: J. Umenberger (email: j.umenberger@acfr.usyd.edu.au).}}

%



\IEEEtitleabstractindextext{%
\begin{abstract}
Estimation of nonlinear dynamic models from data poses many challenges, including model instability and non-convexity of long-term simulation fidelity.
Recently Lagrangian relaxation has been proposed as a method to approximate simulation fidelity and guarantee stability via semidefinite programming (SDP), however the resulting SDPs have large dimension, limiting their utility in practical problems. In this paper we develop a path-following interior point algorithm that takes advantage of special structure in the problem and reduces computational complexity from cubic to linear growth with the length of the data set. The new algorithm enables empirical comparisons to established methods including Nonlinear ARX, and we demonstrate superior generalization to new data. We also explore the ``regularizing" effect of stability constraints as an alternative to regressor subset selection.
\end{abstract}

}

\maketitle

\IEEEdisplaynontitleabstractindextext

%
\IEEEpeerreviewmaketitle

\section{Introduction}

Estimation of predictive mathematical models from data plays an important role in many areas of engineering and science, in particular when models based on first principles are unavailable or too complex. Algorithms that generate such models are a major topic of research in several fields, including statistical inference, machine learning, and system identification (see, e.g., \cite{hastie2009elements, efron2016computer, Ljung1999}). 

In identification of dynamical systems, linear or nonlinear, the {\em model structure} has a strong influence on tractability of the associated optimization problem. One approach is to model the output of the system as a static function of a truncated history of inputs, as in finite-impulse-response  linear models \cite{Ljung1999} and Wiener or Volterra series nonlinear models \cite{schetzen1980volterra}. Although model fitting is usually straightforward, such models are known to be extremely inefficient when representing resonance, i.e. long-term dependencies between inputs and outputs.
Models incorporating {\em internal memory} (i.e. states) and {\em feedback} offer a more efficient and natural representation for such phenomena. 

In this paper, we consider identification of (linear or nonlinear) state-space models of the form
\begin{align}\label{eq:nonlinExpSys}
x_{t+1} = a(x_t,u_t), \quad
y_t = g(x_t,u_t),
\end{align}
where $x_t$ is an internal state, and $u_t, y_t$ are input and output, respectively. This model class is very flexible and includes nonlinear autoregressive models \cite{Sjoberg1995, billings2013nonlinear}, infinite-impulse-response linear systems \cite{Ljung1999}, Hammerstein and Wiener models \cite{billings1982identification},  and recurrent neural networks \cite{lecun2015deep}.

The downside of internal feedback is a substantial increase in the difficulty of the search for a model. 
For instance, when accurate long-term predictive performance is required, it is often appropriate to fit a model by minimization of \emph{simulation error},
a.k.a. output error.
\begin{definition}[Simulation error]\label{def:se}
Given measurements of inputs $\lbrace\tu_t\rbrace_{t=1}^T$ to and outputs $\lbrace\ty_t\rbrace_{t=1}^T$ from some dynamical system, {simulation error} is defined as
\begin{equation}\label{eq:simErrorCost}
\se\defeq {\sum}_{t=1}^{T}\vert\tilde{y}_t - g(x_t,\tu_t)\vert^2
\end{equation}
where $x_t = a(\dots a(a(\tx_1,\tu_1),\tu_2)\dots,\tu_{t-1})$, 
i.e. the solution of \eqref{eq:nonlinExpSys}, with 
input $\lbrace\tu_t\rbrace_{t=1}^T$
and 
initial conditions 
$x_1=\tx_1$.
\end{definition}
Dependence on the simulated internal state $x_t$ renders $\se$ a non-convex function of the model parameters, complicating the search for the global minimum \cite{Sjoberg1995, Ljung2010, paduart2010identification}.
Ensuring stability of the identified model \eqref{eq:nonlinExpSys} is a further challenge \cite{Ljung2010}.

Existing approaches to identification of state-space models include subspace identification for linear systems \cite{van2012subspace}, the prediction error method \cite{Ljung1999}, initializing the search for nonlinear models with frequency-domain fitting of linear models \cite{paduart2010identification}, maximum-likelihood via the expectation-maximization (EM) algorithm \cite{Schon2011,umenberger2015em}, and Bayesian identification via Markov-chain Monte Carlo (MCMC) \cite{ninness2010bayesian}. 
Nonparametric representations of \eqref{eq:nonlinExpSys} are also possible \cite{georgiev1984nonparametric}; e.g., modeling the functions $a$ and $g$ as (realizations of) Gaussian processes \cite{rasmussen2006gaussian} leads to a so-called Gaussian process state-space model (GP-SSM).
GP-SSMs offer considerable flexibility and a principled framework for handling uncertainty. 
Identification methods for such models include EM \cite{turner2010a}, MCMC \cite{frigola2013} and
sequential Monte Carlo \cite{frigola2014variational,Svensson2017flexible}.
However, none of these methods guarantee globally optimal fits or stability of the identified model, 
though in the linear setting methods for ensuring model stability have been proposed
(e.g. \cite{maciejowski1995guaranteed,van2001identification, lacy2003subspace}).

This paper builds upon \cite{tobenkin2017convex}, which proposed a convex parametrization of nonlinear state-space models with guaranteed stability, 
as well as a family of convex upper bounds on simulation error.
This line of research was initiated in \cite{Megretski2008}, and further developed in
\cite{Bond2010,Tobenkin2010,Manchester2012, tobenkin2014}. A central contribution of \cite{tobenkin2017convex} is construction of a simulation-error bound based on a version of the Lagrangian relaxation (LR) \cite{lemarechal2001lagrangian}, closely related to the S-procedure \cite{polik2007survey}. This bound can be represented as a semidefinite program (SDP), however for practical data-sets the resulting SDP is very large, and the experimental results in \cite{tobenkin2017convex} were all based on the simpler but less-accurate {\em robust identification error} (RIE), first presented in \cite{Tobenkin2010}. Indeed, we will show in this paper that when represented as a standard semidefinite program, the LR approach has computational complexity which is {\em cubic} in the length of the data set, severely limiting its practical utility.

Our main contribution is a specialized algorithm that takes advantage of the structure in the LR optimization to significantly improve computational tractability: scaling of Newton iterates with respect to data-set length is now {\em linear} instead of cubic.
Our contribution can therefore be seen in the context of a growing body of research on specialized solvers for special classes of SDPs appearing in robustness analysis via integral quadratic constraints and the Kalman-Yakubovich-Popov lemma, e.g. \cite{kao2004iqc, vandenberghe2005interior, kao2007new, wallin2008cutting}.
Of more direct relevance to system identification is \cite{liu2009interior}, which develops a custom interior point method (exploiting structure in the Nesterov-Todd equations) for nuclear norm approximation with application to subspace identification.
There is also a close relationship to recent developments in sparsity-exploiting SDP solvers, which we discuss in detail in \mysection\ref{sec:sparse}.

A secondary contribution of this paper, enabled by the development of our specialized algorithm, is to empirically evaluate the performance of the LR method and compare it to established methods of linear and nonlinear system identification. In particular, we explore the apparent \emph{regularizing} effect of the stability constraint and LR.

Regularization refers to the process of constraining or reducing model complexity (in some sense) to prevent over-fitting and to manage the  bias-variance trade-off in statistical modeling \cite{hastie2009elements}. Classical methods such as ridge regression (shrinkage) and subset selection (regressor pruning) have long been applied in nonlinear system identification \cite{Sjoberg1995, johansen1997tikhonov, billings2013nonlinear}. More recently, novel regularization strategies have been developed for identification of linear systems, including nuclear norm regularization for subspace identification (e.g. \cite{liu2009interior}) and kernel methods for impulse-response modeling, surveyed in \cite{pillonetto2014kernel}; c.f., also \cite{pillonetto2011new} for extensions of these kernel methods to nonlinear identification. In this paper, we provide evidence that stability constraints and LR have an effective regularizing effect and seem to eliminate the need for regressor pruning.

The structure of the paper is as follows: 
Section \ref{sec:prelim} introduces notation and the problem statement.
\mysection~\ref{sec:lrse} recaps the convex parametrizations of stable models and convex bounds on simulation introduced in \cite{tobenkin2017convex}.  
Section \ref{sec:specialized} contains the main contribution: the specialized algorithm. 
Section \ref{sec:complexity} demonstrates the algorithm's improved computational complexity over existing methods. Sections \ref{sec:nonlinearspring}, \ref{sec:twotank}, and \ref{sec:subspace} present empirical comparisons to established methods on a number of example problems, and finally Section \ref{sec:conc} offers concluding remarks. 
Preliminary work on specialized algorithms in the linear setting was presented in \cite{umenberger2016scalable}, which employed BFGS \cite[\refsec6]{wright1999numerical}) approximations of the Hessian.

\section{Preliminaries}\label{sec:prelim}

\subsection{Notation}
We use the following notation.  
The cone of real, symmetric nonnegative (positive) definite matrices is denoted by $\mathbb{S}^{n}_{+}$ ($\mathbb{S}^{n}_{++}$). 
The $n\times n$ identity matrix is denoted $I_n$. 
Let $\text{vec}:\mathbb{R}^{m\times n}\mapsto\mathbb{R}^{mn}$ denote the function that stacks the columns of a matrix to produce a column vector, and $\text{mat}:\mathbb{R}^{mn}\mapsto\mathbb{R}^{m\times n}$ for its inverse. 
Let $\text{s2v}:\sym^n\mapsto\real^{n(n+1)/2}$  denote the functions that stacks the columns of an $n\times n$ symmetric matrix, with duplicate entries omitted. 
The Kronecker product is denoted $\otimes$. 
The transpose of a matrix $a$ is denoted $a'$, and $|a|_Q^2$ is shorthand for $a'Qa$. 
For a polynomial $p$, $p\in\text{SOS}$ denotes membership in the cone of sum-of-squares polynomials \cite{parrilo2003semidefinite}.

\subsection{Problem statement}\label{sec:problem}
Given measurements of inputs $\lbrace\tu_t\rbrace_{t=1}^T$ to and outputs $\lbrace\ty_t\rbrace_{t=1}^T$ from some dynamical system, we seek a state-space model of the form \eqref{eq:nonlinExpSys} that minimizes the simulation error, $\se$, c.f. Definition~\ref{def:se}.
Furthermore, we require the identified model to be stable in the following sense:
\begin{definition}[Global incremental $\ell^2$ stability]
The model (\ref{eq:nonlinExpSys}) is said to be \textup{stable} if the sequences $\lbrace \bar y_t-\hat{y}_t\rbrace_{t=1}^{\infty}$ and $\lbrace \bar x_t-\hat{x}_t\rbrace_{t=1}^{\infty}$ are square summable for every two solutions $(\bar{u},\bar{x},\bar{y})$ and $(\hat{u},\hat{x},\hat{y})$ of (\ref{eq:nonlinExpSys}), subject to the same input $\bar{u}=\hat{u}$.
\end{definition}
This strong notion of stability ensures sensible model behavior for inputs not present in the training dataset.

\section{Lagrangian relaxation of linearized simulation error}\label{sec:lrse}

In this section, we recap the approach presented in \cite{tobenkin2017convex} to the problem presented in \mysection~\ref{sec:problem}.
This permits the formulation of the optimization problem solved in \mysection~\ref{sec:specialized}.

\subsection{Convex parametrization of stable models}\label{sec:models}
The first major difficulty posed by the problem of \mysection~\ref{sec:problem} is the requirement that the identified model be stable.
This is challenging,
as the simultaneous search for model parameters and a certificate of stability (e.g., a Lyapunov function) is typically nonconvex.
To circumvent this difficulty, \cite{tobenkin2017convex} (building upon \cite{Tobenkin2010}) introduced an \emph{implicit} representation of \eqref{eq:nonlinExpSys}, given by
\begin{align}\label{eq:nonlinImpSys}
e(x_{t+1}) = f(x_t,u_t),  \quad
y_t = g(x_t,u_t),
\end{align}
where $e:\real^{n_x}\mapsto\real^{n_x}$, $f:\real^{n_x\times n_u}\mapsto\real^{n_x}$ and $g:\real^{n_x\times n_u}\mapsto\real^{n_y}$ 
are multivariate polynomials or trigonometric polynomials, \emph{linearly parametrized} by unknown model parameters $\modvar\in\real^{n_\modvar}$. 
We shall enforce that $e(\cdot)$ be a bijection; i.e. for any $b\in\real^{n_x}$ there exists a unique solution $s\in\real^{n_x}$ to $e(s)=b$. 
This means that a model of the form (\ref{eq:nonlinExpSys}) can be recovered by computing $x_{t+1}=e^{-1}(f(x_t,u_t))=a(x_t,u_t)$.
Such a model is said to be \emph{well-posed}.

This implicit representation permits the definition of a convex parametrization of stable models.
Let $\stblset$ denote the set of all models $\modvar$ of the form \eqref{eq:nonlinImpSys} for which $\exists$ $P\in\mathbb{S}^{n_x}_{++}$ and $\postol>0$ such that the matrix inequality 
\begin{align}\label{eq:contractionCondition}
m(\modvar,P,x,u)\defeq& F(x,u)P^{-1}F(x,u)-E(x)-E(x)'+ \nonumber \\ & P + \postol I_{n_x} +G(x,u)'G(x,u)\mleq 0
\end{align}
holds for all $x\in\real^{n_x}$, $u\in\real^{n_u}$ where $E(x)=\nabla_x e(x)$, $F(x,u)=\nabla_x f(x,u)$ and $G(x,u)=\nabla_x g(x,u)$.
The inequality (\ref{eq:contractionCondition}) may be interpreted as a contraction condition \cite{Lohmiller1998} with the metric $E(x)'P^{-1}E(x)$. 
All models 
$\modvar\in\stblset$ 
are guaranteed to be globally incrementally $\ell^2$ stable and well-posed (i.e. $e$ is a bijection), c.f. \cite[Theorem 5]{tobenkin2017convex}.  
Note that \eqref{eq:contractionCondition} is convex in $(\modvar,P)$ for fixed $(x,u)$. 
To ensure \eqref{eq:contractionCondition} holds $\forall \ x,u$, a sum-of-squares (SOS) relaxation is presented in \mysection~\ref{sec:sos}.

\subsection{Linearized simulation error}\label{sec:lse}
The second major barrier to solving the problem of \mysection~\ref{sec:problem} is the existence of local minima due to nonconvexity of simulation error.
Rather than minimize $\se$ directly, the approach proposed in \cite{tobenkin2017convex} is to approximate $\se$ via \emph{Lagrangian relaxation} of the \emph{linearized simulation error}, 
defined as follows:
Given an estimated state sequence $\lbrace\tx_t\rbrace_{t=1}^T$, we define the \emph{equation errors}
\begin{align}\label{eq:ee}
\epsilon_t = f(\tx_t,\tu_t)-e(\tx_{t+1}), \
\eta_t = g(\tx_t,\tu_t)-\ty_t,
\end{align}
and Jacobians
$E_t=\nabla_x \ e\ |_{x=\tx_t}$, $F_t = \nabla_x \ f \ |_{x=\tx_t}^{u=\tu_t}$, $G_t = \nabla_x \ g\ |_{x=\tx_t}^{u=\tu_t}$. The linearized simulation error is then given by
\begin{equation}
\lse = {\sum}_{t=1}^{T}|G_t\Delta_t+\eta_t|^2,
\end{equation} 
where $\Delta_t$ satisfies $\Delta_1=0$ and $E_{t+1}\Delta_{t+1}=F_t\Delta_t+\epsilon_t$ for $t=1,\dots,T-1$.
The linearized simulation error $\lse$ quantifies local (i.e. close to $\lbrace\tx_t\rbrace_{t=1}^T$) sensitivity of the model equations to equation errors; c.f., \cite[\S V]{tobenkin2017convex} for further details.

\subsection{Lagrangian relaxation}\label{sec:lr}
In this work, Lagrangian relaxation refers to the approximation of the nonconvex problem $\min_\modvar \lse$ by the convex problem $\min_\modvar\lrlse(\modvar)$, where 
\begin{align}\label{eq:lseBoundExpand}
\lrlse(\modvar) = \sup_\Delta \ \bigg\lbrace &{\sum}_{t=1}^{T}|G_t\Delta_t+\eta_t|^2 - \lambda_1'E_1\Delta_1 \nonumber \\ -&{\sum}_{t=1}^{T-1}\lambda_{t+1}'(E_{t+1}\Delta_{t+1}-F_t\Delta_t-\epsilon_t) \bigg\rbrace.
\end{align} 
Here, $\Delta=[\Delta_1',\dots,\Delta_T']'\in\real^{Tn_x}$. 
$\lrlse(\modvar)$ represents a convex upper bound on $\lse$ for arbitrary multipliers $\lambda_t$. 
In this paper, we will use 
$\lambda_t=2\Delta_t$, 
due to the desirable properties outlined in the following theorem:

\begin{theorem}[{\cite[Theorem 6]{tobenkin2017convex}}]
For any arbitrary dataset $\tilde{z}\defeq\lbrace\tu_t,\ty_t,\tx_t\rbrace_{t=1}^T$ and $\lambda_t=2\Delta_t$, $0\leq\lrlse(\modvar)<\infty$ for all $\modvar\in\stblset$.
Furthermore, if $\tilde{z}$ represents noiseless inputs, outputs and states from some true model $\modvar^*\in\stblset$, then $\min_{\modvar\in\stblset}\lrlse(\modvar)=0$. 
\end{theorem}

The Lagrangian relaxation \eqref{eq:lseBoundExpand} depends on a \emph{surrogate} state sequence $\lbrace \tx_t\rbrace_{t=1}^T$, c.f. \eqref{eq:ee}. While it is not assumed that these are {\em true} internal states, the more accurate they are the more effective our approach will be. Methods for generating state estimates from input-output data include  subspace methods for linear systems \cite{van2012subspace}.
For nonlinear systems, state estimation is more challenging and solutions can be quite case specific. 
Possible strategies include: subspace methods in the case of weakly nonlinear systems, c.f. Section \ref{sec:twotank}; exploiting physical or structural knowledge, c.f. Section \ref{sec:nonlinearspring}; 
alternating between model-based state estimation and model refinement, e.g. expectation-maximization \cite{umenberger2015em}; 
and using truncated histories of inputs and outputs, as in nonlinear ARX \cite{Sjoberg1995}.  
 
\par For what follows, it is convenient to introduce the following `lifted' representation of \eqref{eq:lseBoundExpand}. Let  
$\mathcal{G}(\liftvar)=\textup{blkdiag}(G_1,\dots,G_T)$, 
$\eta(\liftvar)=\left[\eta_1',\dots,\eta_{T}' \right]'$, 
$\epsilon(\liftvar)=\left[0,\epsilon_1',\dots,\epsilon_{T-1}'\right]'$ and
\begin{align}\label{eq:lifted}
\mathcal{F}(\liftvar)&=\left[\begin{array}{cccc}
E_1 & 0 & 0 & \dots \\
-F_1 & E_2 & 0 & \ddots \\
0 & -F_2 & E_3 & \ddots \\
\vdots & \ddots & \ddots & \ddots
\end{array}\right]. 
\end{align} 
The upper bound in (\ref{eq:lseBoundExpand}) can then be more compactly expressed as 
$\lrlse(\modvar)=\sup_\Delta\llse(\modvar,\Delta)$, where
\begin{equation}\label{eq:lagrangian}
\llse(\liftvar,\Delta)=|\mathcal{G}(\liftvar)\Delta+\eta(\liftvar)|^2 - 2\Delta'(\mathcal{F}(\liftvar)\Delta - \epsilon(\liftvar)).
\end{equation}

\subsection{Optimization with general-purpose solvers}\label{sec:gps}
Minimization of $\lrlse(\modvar)$ can be formulated as the following SDP, compatible with any general-purpose SDP solver:
\begin{subequations}\label{eq:sdp}
\begin{align}
\min_\modvar \quad & s \\
\textup{s.t.} \quad &\left[\begin{array}{ccc}
s & \epsilon(\modvar)' & \eta(\modvar)' \\
\epsilon(\modvar) & \mathcal{F}(\modvar)+\mathcal{F}(\modvar)' & \mathcal{G}(\modvar)' \\
\eta(\modvar) & \mathcal{G}(\modvar) & I_{Tn_y}
\end{array}\right]\mgeq 0 \label{eq:sdp_lmi}
\end{align}
\end{subequations}
where $s$ is a slack variable.
If no structural properties (e.g. sparsity) of (\ref{eq:sdp_lmi}) are exploited by the solver, then each iteration of a primal-dual interior point method requires 
\[O\left(\max\lbrace n_\modvar(n_x+n_y)^3T^3, \ n_\modvar^2(n_x+n_y)^2T^2\rbrace\right)\]
operations to solve,
c.f., e.g., \cite[\S2]{liu2009interior},
where $n_\modvar$ is the number of free model parameters, and $T$ is the number of data points in the training set. In a typical system identification scenario, the model and parameter dimensions $n_x, n_y$, and $n_\rho$ remain moderate in size while the data set length $T$ may be very large. This implies $O(T^3)$ complexity, which will be demonstrated empirically in \mysection\ref{sec:analysis}.  

\section{Specialized algorithm}\label{sec:specialized}

In this section we present the main contribution of this paper: an efficient, scalable algorithm for 
the problem
$\min_{\modvar\in\stblset}\lrlse(\modvar)$,
where $\lrlse(\modvar)$ (c.f. (\ref{eq:lseBoundExpand})) is the convex upper bound \eqref{eq:lseBoundExpand} on linearized simulation error, and $\stblset$ is the convex parametrization \eqref{eq:nonlinImpSys}, \eqref{eq:contractionCondition} of stable models.
Specifically, we present an interior-point algorithm for which the complexity of each Newton iteration grows \emph{linearly} with the number of data points, $T$. 
See Algorithm \ref{alg:ipm} for a complete listing.

\subsection{Explicit LMI representation of stable models}\label{sec:sos}
As discussed in \mysection~\ref{sec:models}, 
the convex set $\stblset$ of stable models is defined by an infinite family of matrix inequalities, i.e., \eqref{eq:contractionCondition}.  
In what follows, we derive an explicit linear matrix inequality (LMI) approximation of $\stblset$, based on sum-of-squares (SOS) programming \cite{parrilo2003semidefinite}.
By the Schur complement, \eqref{eq:contractionCondition} is equivalent to the infinite family of LMIs:
\begin{align}\label{eq:lmiContractionCond}
& M(\modvar,P,x,u) \defeq \\
& \left[\begin{array}{ccc}
E(x) + E(x)' - P - \postol I_{n_x} & F(x,u)' & G(x,u)' \\
F(x,u) & P & 0 \\
G(x,u) & 0 & I_{n_y}
\end{array}\right]\mgeq 0. \nonumber
\end{align}
Introducing $v\in\real^{2n_x+n_y}$, and 
$z=[x',u',v']'\in\real^{n_z}$, 
we define the linearly parametrized scalar polynomial 
\begin{equation}\label{eq:psos}
\psos(z)\defeq v'M(\modvar,P,x,u)v. 
\end{equation}
Then the condition $M(\modvar,P,x,u)\mgeq0 \ \forall \ x,u$ is equivalent to $\psos(z)\geq0 \ \forall z$. 
Testing non-negativity of a general multivariate polynomial is known to be NP-hard. However, constraining $\psos(z)$ to be  SOS  gives tractable \emph{sufficient conditions} for nonnegativity \cite{parrilo2003semidefinite}.
A SOS representation of $\psos(z)$ has the form
\begin{equation}\label{eq:gramPoly}
v'M(\modvar,P,x,u)v=\monos(z)'Q\monos(z) =: {\sum}_{i=1}^{n_q}c_{i}(Q)z^{\beta_i},
\end{equation}
where
$\monos:\real^{n_z}\mapsto\real^{n_\monos}$ 
is a vector of $n_\monos$ monomials and 
$Q\in \sym{^{n_\monos}_+}$  
is the Gram matrix.  
Careful selection of the basis monomials $\monos$ can simplify the SOS program, e.g., reduce the number of constraints and decision variables. 
Tools such as the Newton polytope \cite{Lofberg2009}, and facial reduction \cite{permenter2014basis}, can be used to generate an effective basis. For the examples in this paper, we used the toolbox \cite{spot} for monomial selection.

The SOS representation in \eqref{eq:gramPoly} comprises two constraints:
(i) linear equality constraints such that the coefficients $c_i(Q)$ match those of $\psos$ in \eqref{eq:psos}, and
(ii) nonnegativity of the Gram matrix, $Q\mgeq0$.
Introducing
$\fullvar=[\modvar',\mtovec{P}',\mtovec{Q}']'$, 
the linear equality constraints can be expressed as $A_e\fullvar=b_e$, c.f. Example~\ref{ex:sos}.
Similarly, nonnegativity of the Gram matrix can be encoded as $S(\fullvar)\defeq \textup{mat}(\asos\fullvar)= Q\mgeq0$, where $\asos=[0 \ I]$. 
Then $\psos(z)\in\text{SOS}$ is equivalent to $\fullvar\in\sosset$, where
\begin{equation}\label{eq:sosset}
\sosset=\lbrace \fullvar: \sosmat(\fullvar)=\text{mat}\left(\asos\fullvar\right)\mgeq 0, \ A_e\fullvar=b_e \rbrace.
\end{equation}

\begin{example}\label{ex:sos}
Consider a model \eqref{eq:nonlinImpSys} of the form,
\begin{equation*}
e(x) = \modvar_1x + \modvar_2x^3, \quad f(x,u) = \modvar_3x+u, \quad g(x,u) = x.
\end{equation*}
The polynomial $\psos$ defined in \eqref{eq:psos} is then given by
\begin{equation*}
\psos(z) = (2\rho_1-P-\postol)z_1^2 + 6\modvar_2x^2z_1^2+2\modvar_3z_1z_2 + Pz_2^2 + 2z_1z_3 + z_3^2. 
\end{equation*}
A suitable monomial basis for $\psos(z)$ is $\monos=[z_1x,z_1,z_2,z_3]'$.
The Gram matrix takes the form $Q\in\sym_{+}^4$, which gives 
\begin{equation*}
\theta = \left[\modvar_1 \ \modvar_2 \ \modvar_3 \ P \ Q_{11} \ Q_{12} \dots Q_{44}\right].
\end{equation*}
The equality constraints to ensure $\monos'Q\monos=\psos(z)$ are: $Q_{11}=6\modvar_2$, $Q_{22}=2\modvar_1-P-\postol$, $Q_{33}=P$, $Q_{44}=1$, $Q_{14}=2$, $Q_{23}=2\modvar_3$, $Q_{ij}=0$ for all other $i,j$.
Each of these equality constraints corresponds to one row of $A_e$ and $b_e$. 
For instance, $Q_{11}-6\modvar_2=0$ corresponds to  
\begin{align*}
A_e(1,:) = [0, \underbrace{-6}_{\modvar_2}, 0, \dots, 0, \underbrace{1}_{Q_{11}},0,\dots,0], \ b_e(1) = 0,
\end{align*}
the constraint $Q_{22}-2\modvar_1+P=-\postol$ corresponds to
\begin{align*}
A_e(2,:) = [\underbrace{-2}_{\modvar_1},0,0,\underbrace{-1}_{P},0,\dots,\underbrace{1}_{Q_{22}},\dots,0], \ b_e(2) = -\postol,
\end{align*}
and so on.
\end{example}

\subsection{Structural properties of Lagrangian relaxation}\label{sec:structure}
Given the convex parametrization of stable models $\sosset$, c.f. \eqref{eq:sosset}, our problem becomes $\min_{\fullvar\in\sosset} \ \lrlse(\fullvar)$.
With some abuse of notation, we write $\lrlse(\fullvar)$ in place of $\lrlse(\modvar)$.
Recall from (\ref{eq:lseBoundExpand}) that $\lrlse(\fullvar)\defeq\sup_\Delta\llse(\fullvar,\Delta)$, where $\llse$ is the Lagrangian defined in (\ref{eq:lagrangian}).
$\llse$ is quadratic in $\Delta$, and can be expressed as
\begin{equation}
\llse(\fullvar,\Delta) = \Delta'\ddM\Delta + 2\ddm'\Delta,
\end{equation}
where 
$\ddM\defeq \mathcal{G}'\mathcal{G} - \mathcal{F} - \mathcal{F}'$ and
$\ddm \defeq -\mathcal{G}'\eta-\epsilon$.
The supremum $\sup_\Delta\llse(\fullvar,\Delta)$ is finite if and only if $\llse$ is concave, i.e., $\ddM\mleq0$. 
Imposing strict negative-definiteness ensures robustness and a unique maximizing $\Delta$. 
It turns out that 
$\fullvar\in\sosset$
is sufficient to guarantee $\ddM\ml 0$.
Specifically, we have the following result: 
\begin{prpty}[{\cite[Theorem 6]{tobenkin2017convex}}]\label{prp:finiteSup}
$\fullvar\in\sosset$ implies 
$\ddM\defeq \mathcal{G}'\mathcal{G} - \mathcal{F} - \mathcal{F}' \ml 0$, 
i.e., $\lrlse(\fullvar)$ is finite.
\end{prpty}
The key point is that we can guarantee $\ddM\ml 0$ (a large LMI that grows linearly in dimension with $T$) by enforcing 
$\fullvar\in\sosset$
(a convex constraint that does not grow with $T$).
When $\ddM\ml0$, we have $\lrlse(\fullvar)=\llse(\fullvar,\Delta^*(\fullvar))$, where 
\begin{align}\label{eq:maxDelta}
\Delta^*(\fullvar) = \arg\max_\Delta \ \llse(\fullvar,\Delta) 
 = -\ddM^{-1}\ddm
\end{align}
is the unique maximizing $\Delta$.
By the chain rule
\begin{equation}\label{eq:dldt}
\frac{\p\lrlse}{\p\fullvar}=\frac{\p\llse}{\p\fullvar} + \frac{\p\llse}{\p\Delta}\frac{\p\Delta^*}{\p\fullvar}.
\end{equation}
\begin{prpty}\label{prp:simple}
The gradient of $\lrlse(\fullvar)=\llse(\fullvar,\Delta^*(\fullvar))$ does not depend on 
$\frac{\p\Delta^*}{\p\fullvar}$, nor does the Hessian depend on $\frac{\p^2\Delta^*}{\p\fullvar^2}$.
\end{prpty}
To see this, consider the gradient of $\lrlse(\fullvar)$ w.r.t. $\fullvar$ at a particular parameter $\fullvar^\dagger\in\sosset$.  
As $\Delta^*(\fullvar^\dagger)$ maximizes the smooth function $\llse(\fullvar^\dagger,\Delta)$, we have $\frac{\p \llse}{\p\Delta}=0$ at $\Delta=\Delta^*(\fullvar^\dagger)$, and so \eqref{eq:dldt} reduces to
\begin{equation}\label{eq:gradientResult}
\frac{\p\lrlse}{\p\fullvar} = \left.\frac{\p \llse}{\p\fullvar}\right|_{\fullvar=\fullvar^\dagger, \Delta=\Delta^*(\fullvar^\dagger)}. 
\end{equation}
The key point is that $\frac{\p\Delta^*}{\p\fullvar}$
need not be computed to calculate the gradient of $\lrlse(\fullvar^\dagger)$, which is given by
\begin{equation}\label{eq:gradientFormula}
\frac{\p\lrlse}{\p\fullvar(i)}=2(\mathcal{G}\Delta^*+\eta)'(\mathcal{G}_i\Delta^*+\eta_i) - 2\Delta^{*'}(\mathcal{F}_i\Delta^*-\epsilon_i),
\end{equation}
where $\mathcal{G}_i,\eta_i,\mathcal{F}_i,\epsilon_i$ denote $\sfrac{\p\mathcal{G}}{\p\theta(i)}, \sfrac{\p\eta}{\p\theta(i)}, \sfrac{\p\mathcal{F}}{\p\theta(i)}, \sfrac{\p\epsilon}{\p\theta(i)}$, respectively. 
The Hessian $\nabla^2\lrlse$ is given by
\begin{align}\label{eq:exactHessian}
\frac{\p^2\lrlse}{\p\fullvar(j)\p\fullvar(i)} = &
\frac{\p^2\llse}{\p\fullvar(j)\p\fullvar(i)} + 
\frac{\p^2\llse}{\p\Delta\p\fullvar(i)}\frac{\p\Delta^*}{\p\fullvar(j)} \\ & +
\frac{\p^2\llse}{\p\Delta\p\fullvar(j)}'\frac{\p\Delta^*}{\p\fullvar(i)}  +
\frac{\p\Delta^*}{\p\fullvar(j)}'\frac{\p^2\llse}{\p\Delta^2}\frac{\p\Delta^*}{\p\fullvar(i)}. \nonumber
\end{align}
Notice that $\frac{\p^2\Delta^*}{\p\fullvar^2}$ does not appear in (\ref{eq:exactHessian}), 
for the same reason that $\frac{\p\Delta^*}{\p\fullvar}$ does not appear in 
$\nabla\lrlse$, namely: 
because $\frac{\p\llse}{\p\Delta}(\fullvar,\Delta^*(\fullvar))=0$
for all $\fullvar\in\sosset$.
Specifically, we have
\begin{subequations}\label{eq:hessian_parts}
\begin{align}
\frac{\p^2\llse}{\p\fullvar(j)\p\fullvar(i)}&=2(\mathcal{G}_j\Delta^*+\eta_j)'(\mathcal{G}_i\Delta^*+\eta_i), \label{eq:hess_1} \\
\frac{\p^2\llse}{\p\Delta\p\fullvar(i)} &= \begin{array}{l}
2\Delta^{*'}\left( \mathcal{G}'\mathcal{G}_i + \mathcal{G}_i'\mathcal{G} - \mathcal{F}_i'-\mathcal{F}_i\right) + \\ 2\left(\eta'\mathcal{G}_i+\eta_i'\mathcal{G}+\epsilon_i'\right), \label{eq:hess_2}
\end{array}
\end{align} 
\end{subequations}
and $\frac{\p^2\llse}{\p\Delta^2}=2\ddM$.
To compute $\frac{\p\Delta^*}{\p\fullvar}$ 
rewrite (\ref{eq:maxDelta}) as 
\begin{equation}\label{eq:altMaxDelta}
\ddM\Delta^*=\ddm.
\end{equation}
Application of the product rule to (\ref{eq:altMaxDelta}) yields 
\begin{equation}\label{eq:dDelta}
\ddM\frac{\p\Delta^*}{\p\fullvar(i)}=\left(\frac{\p\ddm}{\p\fullvar(i)}-\frac{\p\ddM}{\p\fullvar(i)}\Delta^*\right),
\end{equation}
from which we can solve for $\frac{\p\Delta^*}{\p\fullvar(i)}\in\real^{Tn_x}$.

To compute $\Delta^*$ and $\frac{\p\Delta^*}{\p\fullvar(i)}$ we need to solve the linear systems \eqref{eq:altMaxDelta} and \eqref{eq:dDelta}, resp., both of which require $\ddM^{-1}$.

\begin{prpty}\label{prp:diag}
$\ddM\defeq \mathcal{G}'\mathcal{G} - \mathcal{F} - \mathcal{F}'$ is block-Toeplitz, Hermitian, and negative definite.
\end{prpty}
Based on these properties of $\ddM$, we can employ the block Thomas algorithm \cite[\refsec 3.8.3]{quarteroni2010numerical},
to 
solve \eqref{eq:altMaxDelta} and \eqref{eq:dDelta} with $O(T)$ operations.


\subsection{Path-following interior point method}\label{sec:pathfollowing}
The algorithm we propose solves
$\min_{\fullvar\in\sosset} \lrlse(\fullvar)$
via a (primal-only) path following interior point, or \emph{barrier}, method; see, e.g., \cite{nesterov1994interior}.
Primal-dual interior point methods are generally expected to be more efficient than barrier methods on standard SDPs \cite{vandenberghe1996semidefinite}.
Despite this, we employ a primal-only method for the following reasons.
Foremost, 
the LR approach requires minimization of a smooth {\em nonlinear} function of the semidefinite cone. 
Unlike standard SDPs, the dual function does not have a simple explicit representation. 
Lifting to a standard-form SDP involves introducing a large number of additional variables (see Section \ref{sec:gps}). 
Second,
using a barrier method, model stability is guaranteed at each iteration.
This permits early stopping (without compromising model stability), a well-known regularization method that has long been used in system identification, c.f., e.g., \cite{Sjoberg1995}.
For standard primal-dual methods, the iterates are not necessarily feasible, except in the limit as the algorithm converges \cite[\refsec11.7.2]{Boyd2004}.
Finally,
as we will see in \mysection\ref{sec:numerical} there appears to be no loss in accuracy associated with our primal-only barrier method compared to primal-dual methods, due to the numerical problems encountered by general-purpose solvers for large datasets.


In developing our primal-only barrier method, we choose to eliminate the equality constraints $A_e\fullvar=b_e$ in $\sosset$, c.f. \eqref{eq:sosset}. 
This is achieved by constructing a general solution to $A_e\fullvar=b_e$, parametrized by $\decvar$, given by
\begin{equation}\label{eq:fullvarparam}
\fullvar(\decvar)=\fullvar^*+N_e\decvar.
\end{equation}
Here $\fullvar^*$ is a particular solution satisfying $A_e\fullvar^*=b_e$, $N_e$ is a basis for the nullspace of $A_e$, and $\decvar$ denotes our new decision variables. The particular solution $\fullvar^*$ can be obtained, e.g., from the semidefinite feasibility problem: 
\begin{align}\label{eq:initprob}
\sosmat(\fullvar)=\text{mat}\left(A_s\fullvar\right)\mgeq 0 , \ A_e\fullvar =b_e.
\end{align}
With the parameterization (\ref{eq:fullvarparam}) the model set $\sosset$ reduces from $\lbrace \fullvar: S(\fullvar)\mgeq 0, A_e\fullvar=b_e\rbrace$ to $\lbrace \decvar:S(\fullvar(\decvar))\mgeq 0\rbrace$, with $S(\cdot)$ defined in \eqref{eq:sosset}.
Our optimization problem then becomes
\begin{equation}\label{eq:problem}
\min_\decvar \ \lrlse(\decvar) \ \text{s.t. } S(\decvar)\mgeq 0.
\end{equation} 
Here, 
we have used $\lrlse(\decvar)$ as shorthand for $\lrlse(\fullvar(\decvar))$.  
Similarly, $S(\decvar)$ is shorthand for $S(\fullvar(\decvar))$. 


The key idea in a path-following interior point method is the introduction a barrier function that tends towards infinity at the boundary of the feasible set. 
We use the standard choice \cite{nesterov1994interior} for the LMI constraint $\sosmat(\decvar)\mgeq 0$, i.e.,
\begin{equation*}
\phi(\decvar)= \begin{cases} 
      -\log\det \sosmat(\decvar) & \sosmat(\decvar)\mg 0 \\
      \quad\infty & \sosmat(\decvar) \mng 0
   \end{cases}. 
\end{equation*}
The barrier function, weighted by a scalar $\tau$, is then added to the objective $\lrlse(\decvar)$ and we solve (using a damped Newton method) a sequence of \textit{unconstrained} optimization problems
\begin{equation}\label{eq:minipm}
\min_\decvar \ \lbrace f_\tau(\decvar) \defeq \lrlse(\decvar) + \tau\phi(\decvar)\rbrace
\end{equation}
for decreasing $\tau$.

\subsection{Newton step}
Given the simplified computation of $\nabla\lrlse$ and $\nabla^2\lrlse$ in \eqref{eq:gradientResult} and \eqref{eq:exactHessian} resp., c.f. Property~\ref{prp:simple}, 
each Newton step for the solution of \eqref{eq:minipm} is entirely standard.
We provide the details here for completeness; c.f., also Algorithm~\ref{alg:ipm}.
   
The gradient of $f_{\tau}(\decvar)$ w.r.t $\decvar$ is given by 
\begin{equation}\label{eq:grad_f}
\nabla f_\tau(\decvar)=\nabla\lrlse(\decvar)+\tau\nabla\phi(\decvar).
\end{equation}
Recalling our parametrization of $\fullvar(\decvar)$ in (\ref{eq:fullvarparam}), we have
\begin{equation}\label{eq:dJ_chain}
\frac{\p \lrlse(\fullvar(\decvar))}{\p \decvar} = \frac{\p \lrlse}{\p\fullvar}\frac{\p\fullvar}{\p\decvar} = \frac{\p \lrlse}{\p\fullvar}N_e
\end{equation}
by the chain rule. 
$\frac{\p \lrlse}{\p\fullvar}$ is given by \eqref{eq:gradientResult}.
Similarly, for the barrier function, the chain rule gives
\begin{equation}\label{eq:dphi_chain}
\frac{\p \phi(\fullvar(\decvar))}{\p \decvar} = \frac{\p \phi(\fullvar)}{\p\fullvar}\frac{\p\fullvar}{\p\decvar} = \frac{\p \phi(\fullvar)}{\p\fullvar}N_e.
\end{equation}
The gradient of $\phi(\fullvar)$ w.r.t. $\fullvar$ is straightforward to compute, as $\sosmat(\fullvar)$ is affine in $\fullvar$; specifically, $\sosmat(\fullvar)=\text{mat}(\asos\fullvar)$. Recall that for $g(Z)=\log\det Z$, where $Z\in\mathbb{S}_{++}$, we have $\nabla g =Z^{-1}$, and so by the chain rule we have
\begin{equation}\label{eq:barGradFormula}
\frac{\p\phi(\fullvar)}{\p\fullvar} = \left[\frac{\p\phi}{\p\fullvar(1)},\dots,\frac{\p\phi}{\p\fullvar({n_\fullvar})}\right] = -\text{vec}(\sosmat(\fullvar)^{-1})'\asos.
\end{equation}

The Hessian of $f_{\tau}(\decvar)$ w.r.t $\decvar$ is given by 
\begin{equation}\label{eq:hess_f}
\nabla^2 f_\tau(\decvar)=\nabla^2\lrlse(\decvar)+\tau\nabla^2\phi(\decvar).
\end{equation}
By the chain rule we have
\begin{equation}\label{eq:d2J_chain}
\frac{\p^2\lrlse(\fullvar(\decvar))}{\p\decvar^2}=\frac{\p\fullvar}{\p\decvar}'\frac{\p^2\lrlse(\fullvar)}{\p\fullvar^2}\frac{\p\fullvar}{\p\decvar} = N_e'\frac{\p^2\lrlse(\fullvar)}{\p\fullvar^2}N_e,
\end{equation}
where $\frac{\p^2\lrlse(\fullvar)}{\p\fullvar^2}$ is given by \eqref{eq:exactHessian}.
The Hessian of the barrier, 
$\nabla^2\phi(\decvar)$ is, like the gradient, straightforward to compute. 
By the chain rule, we have
\begin{equation}\label{eq:d2phi_chain}
\frac{\p^2\phi(\fullvar(\decvar))}{\p\decvar^2}=\frac{\p\fullvar}{\p\decvar}'\frac{\p^2\phi(\fullvar)}{\p\fullvar^2}\frac{\p\fullvar}{\p\decvar} = N_e'\frac{\p^2\phi(\fullvar)}{\p\fullvar^2}N_e.
\end{equation}
While $\nabla^2\phi(\fullvar)$ is easy to compute, it is somewhat cumbersome to express.
Let $\mathcal{B}:\mathbb{S}^{n}\mapsto\mathbb{S}^{n^2}$ denote the function that maps a symmetric matrix $Z\in\mathbb{S}^n$ to the $n\times n$ block matrix, in which the $(i,j)^{\text{th}}$ block is given by $Z(:,j)Z(:,i)'$, where $ \ Z(:,i)$ denotes the $i^{\text{th}}$ column of $Z$. Then, by the chain rule, the Hessian of the barrier function is given by
\begin{equation}\label{eq:barHessFormula}
\nabla^2\phi = \left[\begin{array}{ccc}
\frac{\p^2\phi}{\p\fullvar(1)^2} & \frac{\p^2\phi}{\p\fullvar(1)\p\fullvar(2)} & \dots \\
\vdots & & \ddots
\end{array}\right]=\asos'\mathcal{B}(\sosmat(\fullvar)^{-1})\asos.
\end{equation}
The search direction $d_k$ is then computed in the usual way, c.f. Line~\ref{alg:sd}, and the step length $\alpha$ is selected by a backtracking line search. 

\subsection{Stopping criteria}
For each $\tau$, the `Newton iterations' (L\ref{alg:qn_start}-\ref{alg:qn_end}) terminate when at least one of the following convergence criteria is satisfied: 
i) change in $f_\tau(\decvar)$ is less than a prescribed tolerance, $\delta_f$; 
ii) the maximum absolute value of an element of $\nabla f_\tau(\decvar)$ is less than $\delta_g$; 
iii) the step size $\alpha d_k$ is less than $\delta_f$. 
The `outer iterations' (and thus, the algorithm) terminate when the change in $\lrlse(\decvar)$ is less than a prescribed tolerance, $\delta_J$. 
Recommended values for these parameters are summarized in Table \ref{tab:alg_params}.

\begin{table}
\caption{Parameter values for Algorithm \ref{alg:ipm}.}
\label{tab:alg_params}
\begin{center}
\begin{tabular}[c]{cll}
\hline
Parameter & Description & Value \\\hline
$\tau_0$ & Initial barrier weight & $10^4$ \\
$\beta$  & Barrier weight division factor & 10 \\
$\delta_f$ & Newton objective tolerance & $10^{-10}$ \\
$\delta_g$ & Newton gradient tolerance & $10^{-10}$ \\
$\delta_J$ & Objective convergence tolerance & $10^{-11}$ \\
\texttt{maxit} & Max no. of Newton iterations & $10^4$ \\\hline
\end{tabular}
\end{center}
\end{table}


\begin{algorithm}
\caption{MIN-LAGRANGIAN}\label{alg:ipm}
\begin{algorithmic}[1]
\State Initialize $\fullvar_0=\fullvar^*$, where $\fullvar^*$ is given by (\ref{eq:initprob})
\State Initialize $\decvar=0$
\State Initialize $\tau_0$, c.f. Table \ref{tab:alg_params}
\While{$|\lrlse(\decvar_j)-\lrlse(\decvar_{j-1})|>\delta_J$} \label{alg:obj_converge}
\State $\decvar_k\leftarrow\decvar_j$
\State Set $f_\tau(\decvar)=\lrlse(\decvar)+\tau_j\phi(\decvar)$
\For{$k=1:\texttt{maxit}$}\label{alg:qn_start}
\State Compute $\nabla\lrlse(\decvar_k)$ using (\ref{eq:dJ_chain}) and (\ref{eq:gradientFormula}) \label{alg:dj}
\State Compute $\nabla\phi(\decvar_k)$ using (\ref{eq:dphi_chain}) and (\ref{eq:barGradFormula}) \label{alg:db}
\State Form $\nabla f_\tau(\decvar_k)$ using (\ref{eq:grad_f}) 
\State Compute $\nabla^2\lrlse(\decvar_k)$ using (\ref{eq:d2J_chain}), (\ref{eq:exactHessian}), (\ref{eq:hessian_parts}), (\ref{eq:dDelta}) \label{alg:ddj}
\State Compute $\nabla^2\phi(\decvar_k)$ using (\ref{eq:d2phi_chain}) and (\ref{eq:barHessFormula}) \label{alg:ddb}
\State Form $\nabla^2 f_\tau(\decvar_k)$ using (\ref{eq:hess_f}) 
\State Solve \label{alg:direction} $\nabla^2 f_\tau(\decvar_k)^{-1} d_k=-\nabla f_\tau(\decvar_k)$ \label{alg:sd}
\State Compute the step length $\alpha_k$ by a backtracking line 
\Statex[2] search to satisfy the Wolfe conditions. \label{alg:linesearch}
\State Update the parameter estimate: $\decvar_{k+1}=\decvar_k+\alpha_k d_k$

\IfNoThen {$|f_\tau(\decvar_{k+1})-f_\tau(\decvar_{k})|<\delta_f$ \textbf{ or }} 
\Statex[2] \quad $\Vert\nabla f_\tau(\decvar_{k+1})\Vert_\infty<\delta_g$ \textbf{ or } 
\Statex[2] \quad $\Vert \alpha d_k\Vert_\infty<\delta_f$ \algorithmicthen \label{alg:qn_converge}

\State $\decvar_j\leftarrow\decvar_k$ and \textbf{break}
\EndIf

\EndFor \label{alg:qn_end}
\State Set $\tau_{j+1}=\tau_j/{\beta}$ for some constant $\beta$ \label{alg:tauupdate}
\EndWhile \label{alg:endOuter}
\State\Return $\fullvar = \fullvar_0 + N_e\decvar_j$
\end{algorithmic}
\end{algorithm}

\subsection{Special case: Identification of LTI systems}\label{sec:lti}
We conclude this section by making explicit the ways in which \ours \ is simplified when applied to the special case of LTI systems.
Throughout this section, we use the specific implicit representation of LTI systems
\begin{subequations}\label{eq:impLinSys}
\begin{align}
Ex_{t+1} &=Fx_t+Ku_t, \\
y_t &= Cx_t + Du_t, 
\end{align}
\end{subequations}
where 
$E\in\real^{n_x\times n_x}$, $F\in\real^{n_x\times n_x}$ and $K\in\real^{n_x\times n_u}$.
There are two key simplifications in the linear case.
First, linearized simulation error $\lse$ and simulation error $\se$ are equivalent.
To see this clearly, observe that for linear models $\nabla_x\ e(x) = E$, $\nabla_x\ f(x,u)=F$, $\nabla_x\ g(x,u) = C$, 
$\epsilon_t = F\tilde{x}_t+K\tilde{u}_t - E\tilde{x}_{t+1}$ and
$\eta_t = C\tilde{x}_t+D\tilde{u}_t-\tilde{y}_t$.
Substituting these identities into the definition of $\lse$ with $\Delta_t=x_t-\tx_t$, c.f. \mysection \ref{sec:lr}, we obtain
$
\lse=\sum_{t=1}^{T}|G_t\Delta_t+\eta_t|^2 = \sum_{t=1}^{T}|Cx_t+D\tilde{u}_t - \tilde{y}_t|^2
$
subject to the constraints $\Delta_t=0\iff x_t=\tx_t$ and
$
E_{t+1}\Delta_{t+1} = F_t\Delta_t+\epsilon_t \iff Ex_{t+1}=Fx_t+K\tu_t, 
$
i.e., linearized simulation error $\lse$ equals simulation error $\se$.

\par Second, there is no conservatism in the stability constraint: the stability condition (\ref{eq:lmiContractionCond}) reduces to 
\begin{equation}\label{eq:stabilityCond}
M_l(\modvar,P)=\left[\begin{array}{ccc}
E + E' - P  + \postol I& F' & C' \\
F & P & 0 \\
C & 0 & I
\end{array}\right] \mg 0. 
\end{equation}
As (\ref{eq:stabilityCond}) represents a LMI, there is no need for SOS approximation, as in the nonlinear case.
In fact, $\ltistblset\defeq\lbrace \modvar, P: M_l(\modvar,P)\mg 0\rbrace$ defines a convex parametrization of \emph{all} stable LTI systems, c.f. \cite[Lemma 4]{Manchester2012}, i.e., (\ref{eq:stabilityCond}) is necessary and sufficient for stability.

\section{Computational complexity}\label{sec:complexity}
In this section we examine the computational complexity of the proposed algorithm with respect to the length of the data set $T$. 
We will show that the per-iteration cost of the proposed algorithm grows linearly with $T$, 
a significant improvement over
the $O(T^3)$ per-iteration complexity of general-purpose SDP solvers, c.f. \mysection\ref{sec:gps}. 
This does not result in a {\em complete} complexity analysis, since we do not bound the {\em number} of iterations required. 
However, it is generally observed empirically that the number of iterations required grows very mildly with the number of variables \cite{vandenberghe1996semidefinite}, and we confirm this in the next subsection. In what follows, (L$n$) refers to line $n$ of Algorithm \ref{alg:ipm}.

\subsection{Complexity of each Newton iteration}\label{sec:analysis}
In this subsection we establish that computational complexity of the gradient (L\ref{alg:dj}) and Hessian (L\ref{alg:ddj}) of $\lrlse(\decvar)$ both scale linearly with $T$. 
The gradient (L\ref{alg:db}) and Hessian (L\ref{alg:ddb}) of the barrier function $\phi(\decvar)$, as well as the calculation of the search direction (L\ref{alg:direction}), 
do not depend on $T$. Computation of the gradient $\nabla\lrlse$ requires:
\begin{itemize}
\item one application of the chain rule, as in (\ref{eq:dJ_chain}) which does not grow with $T$, 
\item  $n_\fullvar$ applications of formula (\ref{eq:gradientFormula}). 
Notice, from (\ref{eq:lifted}), that $\mathcal{G}$ and $\mathcal{F}$, along with the derivatives $\mathcal{G}_i$ and $\mathcal{F}_i$, are sparse banded matrices. 
This implies that the products $\mathcal{G}\Delta^*$, $\mathcal{G}_i\Delta^*$ and $\Delta^{*'}\mathcal{F}_i\Delta^*$ in (\ref{eq:gradientFormula}) can be computed with $O(T)$ arithmetic operations. 
As the model (\ref{eq:nonlinImpSys}) is linearly parametrized, the gradients $\mathcal{G}_i,\eta_i,\mathcal{F}_i$ and $\epsilon_i$ can be precomputed off-line.
\item 
The most expensive operation would appear to be the computation of $\Delta^*$ by solving the linear system (\ref{eq:altMaxDelta}).
However, as $\ddM$ is block diagonal, Hermitian, and sign-definite we can employ the block Thomas algorithm \cite[Section 3.8.3]{quarteroni2010numerical}, to compute $\Delta^*$ with $O(T)$ operations.
\end{itemize}
To compute each of the $n_\fullvar(n_\fullvar+1)/2$ unique elements of $\nabla^2\lrlse$, 
we require:
\begin{itemize}
\item one application of (\ref{eq:hess_1}), requiring $O(T)$ operations due to block diagonality of $\mathcal{G}_i$, c.f. (\ref{eq:lifted}),
\item one application of (\ref{eq:hess_2}), requiring $O(T)$ operations due to block diagonality of $\mathcal{G}$, $\mathcal{G}_i$ and $\mathcal{F}_i$, c.f. (\ref{eq:lifted}), 
\item the solution to (\ref{eq:dDelta}) for $i=1,\dots,n_\fullvar$, requiring $O(T)$ operations as $\ddM$ is block tridiagonal, Hermitian and sign-definite, c.f. Property~\ref{prp:diag} in \mysection\ref{sec:structure}.
\item computation of $\frac{\p\Delta^*}{\p\fullvar(i)}'\ddM\frac{\p\Delta^*}{\p\fullvar(i)}$ for $i=1,\dots,n_\fullvar$, requiring $O(T)$ operations, taking $\ddM\frac{\p\Delta^*}{\p\fullvar(i)}$ from (\ref{eq:dDelta}). 
\end{itemize} 
To summarize, the complexity of computing each Newton step of the proposed algorithm is therefore $O(T)$.

Before moving on, we remark that computation of the Hessian is the most expensive part of each iteration.
For identification of `large scale' systems (e.g. models of high dimension $n_x$), it is possible to use only gradient information, if moderate-accuracy is acceptable, e.g., gradient descent or BFGS approximation of the Hessian, as in \cite{umenberger2016scalable}.

\subsection{Convergence behavior}\label{sec:converge}

Barrier methods, such as Algorithm~\ref{alg:ipm}, comprise two nested iterations: 
i) outer iterations (a.k.a. centering steps) in which the barrier weight $\tau$ is decreased, c.f. (L\ref{alg:obj_converge}-\ref{alg:endOuter}), and
ii) inner iterations (a.k.a. Newton steps) by which the centering subproblem $\min_\decvar f_\tau(\decvar)$ is solved, c.f. (L\ref{alg:qn_start}-\ref{alg:qn_end}).
The number of centering steps required for convergence to a specified accuracy is simple to compute (assuming each centering subproblem is solved to sufficient accuracy), c.f. e.g., \cite[\refsec11.3.3]{Boyd2004}.
However, bounding the number of Newton steps \emph{per} outer iteration requires additional assumptions on $f_\tau$, 
namely, that $f_\tau$ be \emph{self-concordant},\footnote{A scalar convex function $f$ is said to be self-concordant if its third derivative is bounded as follows: $|\nabla^3f|\leq2\nabla^2f^{3/2}$, c.f. \cite[\refsec9.6]{Boyd2004}.} 
c.f., e.g., \cite{nesterov1994interior}\cite[\refsec11.5]{Boyd2004}.

In our case it is unclear whether or not $\lrlse$ is self-concordant, and so establishing a bound on the total number of Newton steps for our algorithm is difficult.
Nevertheless, we make the following two remarks.

First, even without self-concordance, convergence of 
each centering subproblem can be guaranteed.
The key is the use of a backtracking line search to choose the step size $\alpha_k$, rather than fixed damping. 
When $\alpha_k$ satisfies the Wolfe conditions (c.f., e.g., \cite[\refsec3.1]{wright1999numerical}), the damped Newton steps $\decvar_{k+1}=\decvar_k+\alpha_k d_k$ converge to $\arg\min_\decvar f_\tau(\decvar)$ as $k\rightarrow\infty$, as long as $d_k$ is a \emph{descent direction}, c.f. \cite[Theorem 3.2]{wright1999numerical}.
To ensure that $d_k$ is a descent direction, it may be necessary to modify the Hessian.
Specifically, one can replace the search direction in (L\ref{alg:sd}) with $d_k=-H_k^{-1}\nabla f_\tau(\decvar_k)$, where $H_k\defeq\nabla^2 f_\tau(\decvar_k)+\regp_k I$.
Here $\regp_k>0$ is chosen such that $H_k$ has bounded condition number, i.e., $\Vert H_k\Vert\Vert H_k^{-1}\Vert\leq C$, for some $C$ and $\forall k$.
Convexity of $\lrlse(\decvar)$ and $\phi(\decvar)$ implies $\nabla^2 f_\tau\mgeq0$, and the addition of $\regp_k I$ (if necessary) ensures that $d_k=-H_k^{-1}\nabla f_\tau$ is a descent direction. 
This strategy of augmenting the Hessian to improve numerical conditioning is often referred to as Hessian modification; c.f. \cite[\refsec3.4]{wright1999numerical}.
In practice, we have found that such Hessian modification was not necessary for convergence; indeed for all the numerical results in this paper we simply take $H_k=\nabla^2 f_\tau(\decvar_k)$.
  
\begin{figure}
\centering
\includegraphics[width=7.5cm]{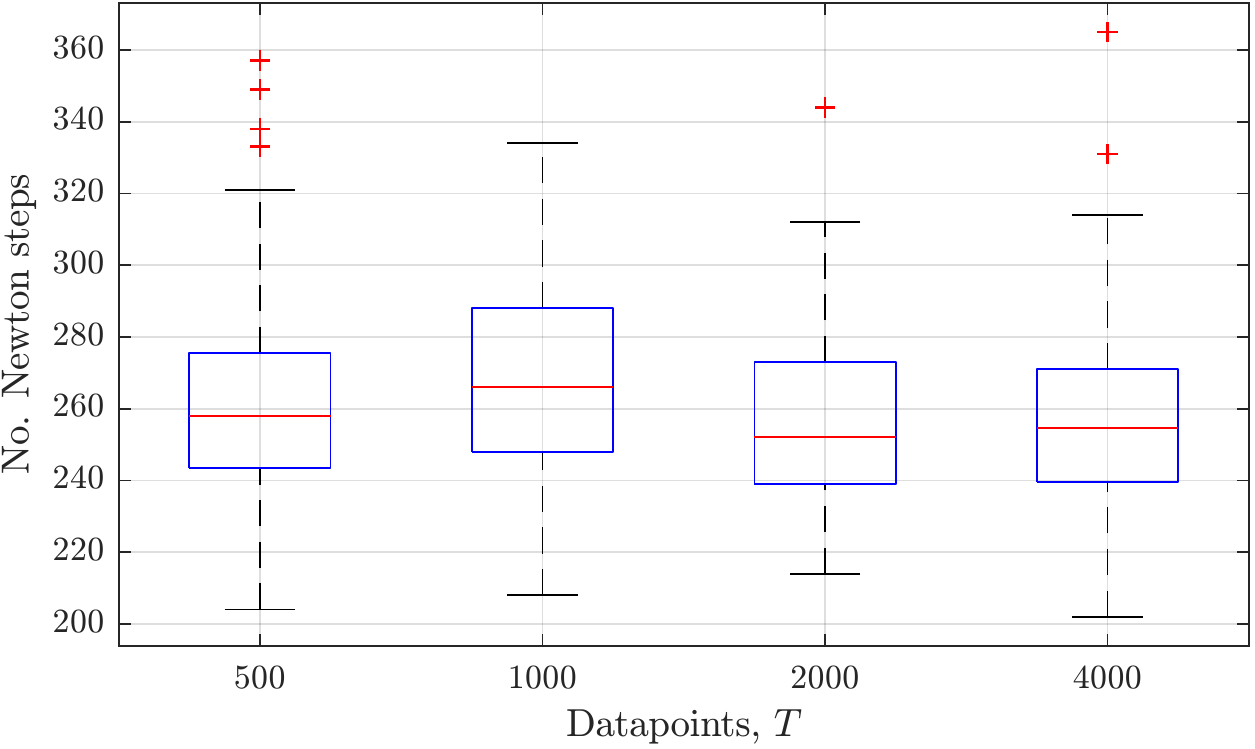}
\caption{
Total number of Newton steps required for identification of a nonlinear model ($n_x=4$) with $(e,f,g)$ of degree (3,3,1). 
On average, the number of Newton steps remains constant with increasing dataset length, $T$. 60 trials were conducted for each $T$.
}
\label{fig:newtons}
\end{figure}

Second, although such a `Hessian modification' ensures convergence of each centering step, without self-concordance it offers no bound on the number of Newton steps required for convergence. 
Nonetheless, we have observed empirically that the number of Newton steps appears to remain constant with increasing dataset length $T$, c.f., e.g., Figure~\ref{fig:newtons}. 
Furthermore, Table~\ref{tab:dj} compares solutions from our specialized algorithm to the primal-dual IPM Mosek.
In all cases (excluding the linear model), our algorithm achieves a better solution (i.e. lower value of $\lrlse(\fullvar)$ with feasible $\fullvar$) than the primal-dual method.
As such, convergence of our primal-only method appears to be reliable, with no loss in accuracy compared to primal-dual methods.  
These empirical results are consistent with the observation that primal-only barrier methods tend to work well on a number of problems for which self-concordance cannot be verified, e.g., geometric programs \cite[\refsec11.5.1]{Boyd2004}.


\begin{table}
\caption{Normalized difference between solutions from our specialized algorithm, $\theta^\textup{s}$, and a primal-dual IPM (Mosek), $\theta^\textup{pd}$, i.e., $(\lrlse(\theta^\textup{pd})-\lrlse(\theta^\textup{s}))/\lrlse(\theta^\textup{pd})$.
Model refers to the degree of the polynomials $(e,f,g)$.
The first column denotes the dataset length $T$.
Five trials were conducted per configuration; the worst (i.e. lowest/most negative) result is recorded.  
}
\label{tab:dj}
\begin{center}
\bgroup
\def\arraystretch{1.2}
\begin{tabular}[c]{l|ccc}
\hline
Model & (1,1,1)  & (3,3,1) & (5,3,1) \\ \hline
200   & $-9.61\tte^{-8}$ & $1.65\tte^{-6}$ & $1.91\tte^{-3}$ \\
300   & $-1.52\tte^{-7}$ & $2.03\tte^{-6}$ & $3.38\tte^{-3}$ \\
400   & $-2.23\tte^{-7}$ & $7.03\tte^{-7}$ & $3.10\tte^{-3}$ \\
500   & $-1.31\tte^{-7}$ & $1.04\tte^{-6}$ & $2.40\tte^{-3}$ \\
1000  & $-4.24\tte^{-8}$ & $8.32\tte^{-7}$ & $4.22\tte^{-3}$ \\
\hline
\end{tabular}
\egroup
\end{center} 
\end{table}

\subsection{Empirical results}\label{sec:numerical}
In this section we provide an empirical comparison between \ours \ and general-purpose solvers. 
Both methods solve the same convex optimization problem, 
$\min_\decvar \ \lrlse(\decvar) \ \text{s.t. } S(\decvar)\geq 0$, as in (\ref{eq:problem}).
All computations were carried out with an Intel i7 (3.40GHz, 8GB RAM). 

\begin{figure}
\centering
\subfloat[Nonlinear identification.]{
\includegraphics[width=8.5cm]{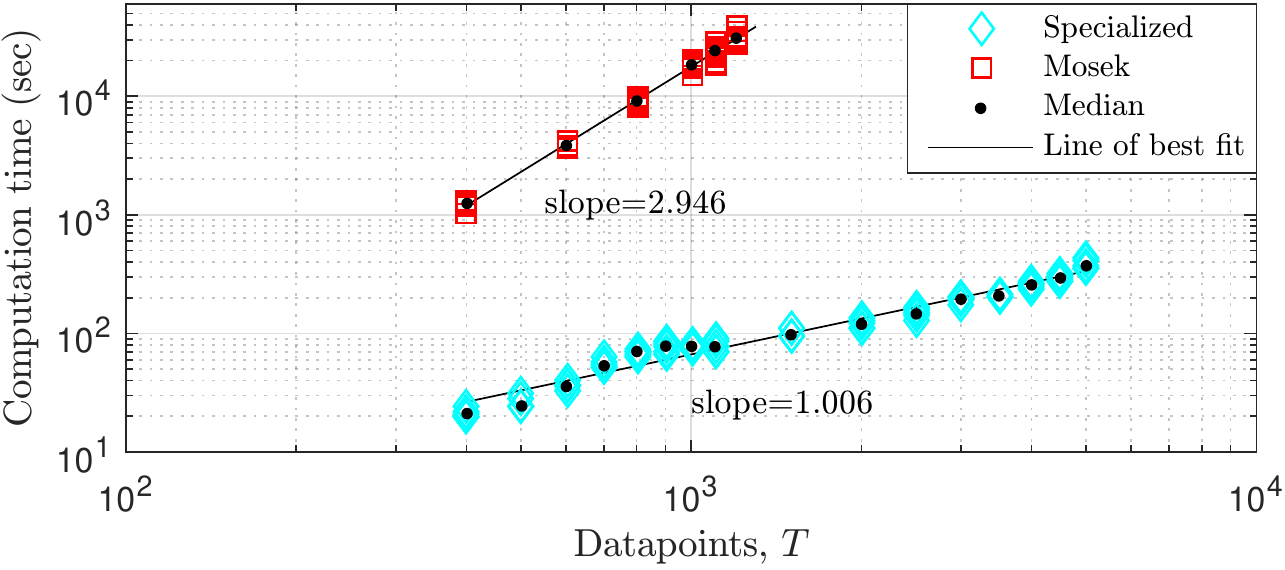}} \\
\subfloat[Linear identification.]{
\includegraphics[width=8.5cm]{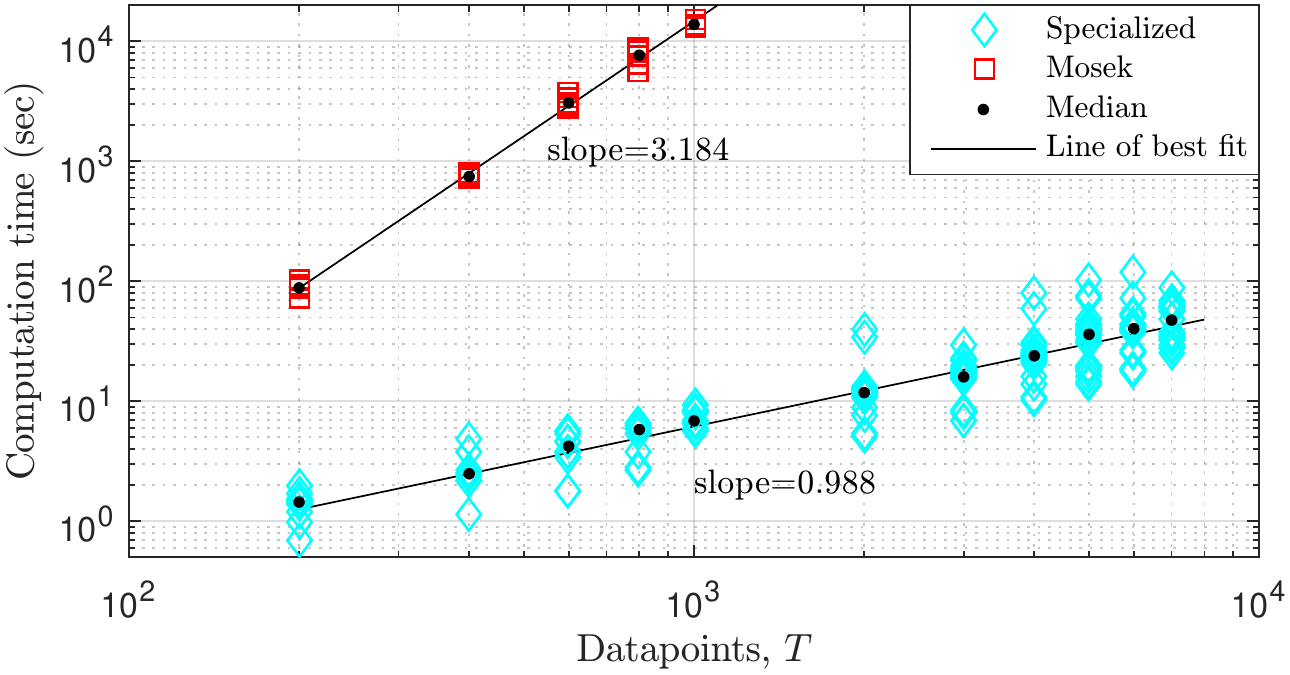}}
\caption{
Computation times for 
solving $\min_\decvar \ \lrlse(\decvar) \ \text{s.t. } S(\decvar)\geq 0$, as in (\ref{eq:problem}),
via two methods: our proposed algorithm (Specialized) and a general-purpose solver (Mosek). 
In (a) the identified SISO nonlinear model is of the form (\ref{eq:lr_model_cs}) with $n_x=4$, $\deg_x(e)=\deg_x(f)=3$ and $\deg_x(g)=1$. For each value of $T$, 10 trials (each with different random noise and input realizations) were conducted.
In (b) the identified $4\myth$ order SISO LTI model is randomly generated for each trial. For each value of $T$, 5 and 20 trials were conducted for Mosek and Specialized, respectively. }
\label{fig:timing}
\end{figure}

We begin with a nonlinear example.
Figure~\ref{fig:timing}(a) 
presents computation times for identification of a SISO nonlinear model of the form (\ref{eq:lr_model_cs}), with $n_x=4$, $\deg_x(e)=\deg_x(f)=3$, and $\deg_x(g)=1$. 
Specifically, we compare our proposed algorithm to Mosek v7.0.0.119 (using Yalmip \cite{lofberg2005yalmip} for SDP formulation), which in our experience is the best currently available general-purpose SDP solver. 
Problem data is generated by simulation of the nonlinear mass-spring-damper depicted in Figure~\ref{fig:msd_1} over time intervals of increasing length $T$. As the focus of this section is algorithmic scalability, we refer the reader to \mysection \ref{sec:msd} for simulation details. 
Examining
Figure~\ref{fig:timing}(a),
it is clear that the specialized algorithm exhibits better scalability compared to Mosek.
In fact, for the specialized algorithm, the slope of the line of best is 1.006 indicating approx. linear growth with $T$, whereas the slope for Mosek is 2.946, indicating approx. cubic growth.
This is consistent with the analysis of \mysection\ref{sec:analysis} and \mysection\ref{sec:gps}.
Furthermore, for $T>1200$, Mosek reports an \cfont{out of memory} error and fails to return a solution.  
At this point it is worth emphasizing that we are comparing a basic Matlab implementation of a primal-only barrier method (i.e., our proposed algorithm) to a highly optimized commercial primal-dual solver (i.e., Mosek).
The superior speed of the former over the latter illustrates the advantages of exploiting problem structure, c.f. \mysection\ref{sec:structure}.
Further improvement in performance (i.e., speed) could likely be achieved with various refinements, e.g., implementation in compiled language, such as \cfont{C/C++}.

Before moving on, we note that in many of the trials depicted in 
Figure~\ref{fig:timing}(a),
Mosek encountered numerical problems, and often reported \cfont{unknown} as the final solution status. 
In all such cases, these solutions turned out to feasible (corresponding to stable models), however, it is not uncommon for primal-dual solvers to return solutions to SOS programs that are (slightly) infeasible, c.f. \cite{Lofberg2009}.
In contrast, our primal only interior-point method ensures feasibility of the solution (i.e. stability of the identified model) at every iteration. 
Furthermore, for every trial depicted in 
Figure~\ref{fig:timing}(a),
the objective value $\lrlse$ attained by \ours \ was lower than the value obtained by Mosek, c.f. also Table~\ref{tab:dj}.

Next, we consider a linear example. 
Figure~\ref{fig:timing}(b) 
presents computation times for identification of $4\myth$ order SISO LTI systems, again comparing \ours \ to Mosek. 
In each trial, the true system was randomly generated using Matlab's \cfont{drss} function, and simulated for $T$ timesteps, excited by a white noise input. The output was corrupted by additive white noise to give a SNR of 17dB, and N4SID \cite{van2012subspace} was used to obtain the state estimates $\lbrace\tx_t\rbrace_{t=1}^T$.
As in the nonlinear example, the results support the claim that scalability of the specialized algorithm linear w.r.t. $T$, while Mosek is cubic, although there is slightly more variability in computation time due to the randomly generated test systems.
Finally, Table \ref{tab:compTime_varOrder} records computation times for varying model order $n_x$, with the length of the dataset held constant at $T=400$ in all trials.


\begin{table}
\caption{Computation time (in seconds, to 3 s.f.) for varying model order $n_x$ (SISO LTI) and $T=400$, averaged over $5$ trials.}
\label{tab:compTime_varOrder}
\begin{center}
\begin{tabular}[c]{l|rrrr}
\hline
Model size, $n_x$ & 2 & 4 & 6 & 8 \\\hline
Specialized algorithm & 0.339 & 2.74 & 8.74 & 34.9 \\
Mosek 7.0.0.119 & 162 & 882 & 2550 & 7340 \\
\hline
\end{tabular}
\end{center} 
\end{table}

\subsection{Relationship to sparsity-exploiting solvers}\label{sec:sparse}

Recall from \mysection\ref{sec:pathfollowing}, one of the main motivations for optimizing $\lrlse$ directly was avoiding the lifted representation (\ref{eq:sdp}) required by general-purpose solvers. 
In this lifted formulation, the dimension of the LMI (\ref{eq:sdp_lmi})
grows linearly with the number of data points, $T$, leading to worst-case per-iteration computational complexity that is cubic in $T$.

Though large, the LMI (\ref{eq:sdp_lmi}) is highly structured.
In fact, as $\mathcal{F}$ and $\mathcal{G}$ are block Toeplitz and block diagonal, respectively, (\ref{eq:sdp_lmi}) has a sparsity pattern characterized by a \emph{chordal} graph.
Since the early 2000s, there has been considerable research into exploiting chordal sparsity in semidefinite programming, c.f. 
\cite{fukuda2001exploiting,nakata2003exploiting,burer2003semidefinite,kim2012exploiting,srijuntongsiri2004fully,vandenberghe2015chordal}.
One such example is the recent work \cite{andersen2010implementation}, which presents nonsymmetric interior-point methods for optimization over semidefinite cones with chordal sparsity patterns.
A number of algorithms for computing the search direction (i.e. solving the Newton equations) in primal scaling and dual scaling methods are derived, based largely on the zero fill-in Cholesky factorization for matrices with chordal sparsity, c.f. \cite[\refsec4]{andersen2010implementation}.  
The authors utilize these algorithms in a feasible-start primal scaling method, and show that per-iteration computational complexity grows linearly
with the LMI dimension.

The solver presented in \cite{andersen2010implementation} and the algorithm we propose in this paper each have per-iteration complexity that scales linearly with data length $T$; although 
\cite{andersen2010implementation} is of course more generally applicable.
The difference is, \cite{andersen2010implementation} exploits chordal sparsity, c.f. Property~\ref{prp:diag}, whereas we exploit structural properties of Lagrangian relaxation, c.f., Properties~\ref{prp:finiteSup}, \ref{prp:simple} and \ref{prp:diag}.

\section{Case study: Mechanical system with nonlinear spring}\label{sec:nonlinearspring}

In this section, we consider identification of a mechanical system with nonlinear spring stiffness. Accurate modeling of such systems is critical in several application areas, e.g. microelectromechanical systems (MEMS) \cite{mestrom2008modelling} and precision motion control \cite{otsuka1998influence}.

\subsection{System Description}\label{sec:msd}
\begin{figure}
\centering
\includegraphics[width=6cm]{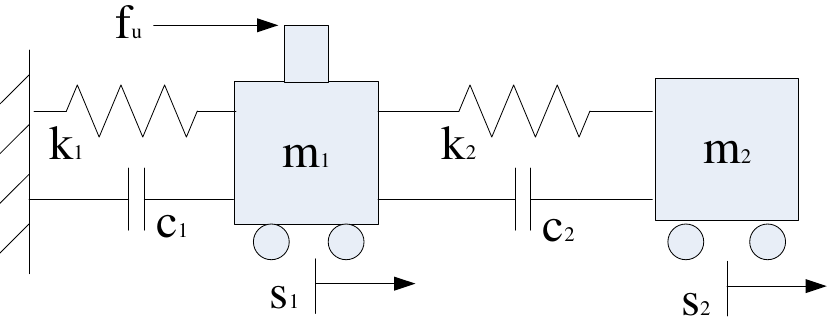}
\caption{Mass-spring-damper system, with parameters $m_1=0.5\ {kg}$, $m_2=0.1\ {kg}$, $c_1=0.01\ Nsm^{-1}$, $c_2=0.1 Nsm^{-1}$. The spring has the nonlinear force-displacement curve (\ref{[eq:nlspring]}) with $k_1=2$, $k_2=1$. The measured control input is force $f_u$ and the measured system outputs are the displacements $s_1$ and $s_2$.}
\label{fig:msd_1}
\end{figure}

A schematic of the system is shown in Figure \ref{fig:msd_1}. The springs have a nonlinear characteristic given by:
\begin{equation}\label{[eq:nlspring]}
k(s) =  k\tan\left(\frac{\pi s}{2\times 1.25}\right), \ s\in[-1.25,1.25]. \\ 
\end{equation}
To generate training data, the system is simulated for 100 seconds with \texttt{ode45}, excited by a superposition of sinusoidal forces, each with randomized frequency, phase and amplitude. 
We sample the input force $\tilde{f}$ and the displacement of the two masses, $s^{(1)}$ and $s^{(2)}$, at 10Hz, to give discrete time data $\tilde{f}_t=\tilde{f}(t\times T_s)$ and $s^{(i)}_t=s^{(i)}(t\times T_s)$, $i=\lbrace1,2\rbrace$, $T_s=0.1$.
We then corrupt the displacement data with additive Gaussian noise
$
\tilde{s}_t^{(i)}=s^{(i)}_t+w_t^{(i)}, \ w_t^{(i)}\sim\mathcal{N}(0,10^{-4}), \ i\in\lbrace 1,2\rbrace,
$
to simulate measurement errors, giving a signal-to-noise ratio (SNR) of approx. 34dB.
Our goal is to model the dynamics from the input force to the position of the second mass, i.e.,
$\lbrace \tilde{u}_t,\tilde{y}_t\rbrace_{t=1}^T = \lbrace \tilde{f}_t,\tilde{s}_t^{(2)}\rbrace_{t=1}^T$
with $T=10^3$.
To estimate the internal states $\lbrace\tilde{x}_t\rbrace_{t=1}^T$, used in the construction of the Lagrange multipliers, c.f. \mysection \ref{sec:lr}, we take
\begin{equation*}
\tilde{x}_t=\left[\tilde{s}_t^{(1)}, \ \tilde{s}_t^{(2)}, \ \frac{\tilde{s}_{t+1}^{(1)}-\tilde{s}_{t-1}^{(1)}}{T_s}, \ \frac{\tilde{s}_{t+1}^{(2)}-\tilde{s}_{t-1}^{(2)}}{T_s} \right]',
\end{equation*}
i.e., we exploit our knowledge of the system structure and approximate the velocities by the central difference.

\par In the following case studies, we will apply Lagrangian relaxation to implicit models of the form (\ref{eq:nonlinImpSys}), with
\begin{subequations}\label{eq:lr_model_cs}
\begin{align}
e: \real^{n_x}\mapsto\real^{n_x} &= [e_1(x),\dots,e_{n_x}(x)]', \\
f: \real^{n_x}\times\real^{n_u}\mapsto\real^{n_x} &= [f_1(x,u),\dots,f_{n_x}(x,u)]', \\
g: \real^{n_x}\times\real^{n_u}\mapsto\real^{n_y} &= [g_1(x,u),\dots,g_{n_y}(x,u)]'.
\end{align}
\end{subequations}
Each function $e_i$, $f_i$ and $g_i$ is a scalar valued, multivariate polynomial, the degree of which will be specified for each application example. 
We will use the term ``degree $n$'', and the notation $\deg_x(p)=n$, to refer to a polynomial $p$ containing all possible monomials in $x$ up to degree $n$, e.g., if $n_x=2$ then ``$e_1$ \emph{is degree} $2$'', or $\deg_x(e_1)=2$, implies
\begin{equation*}
e_1(x) = \modvar_0 + \modvar_1x_1 + \modvar_2x_2 + \modvar_3x_1x_2 + \modvar_4x_1^2 + \modvar_5x_2^2
\end{equation*}
where $\lbrace\modvar_i\rbrace_{i=0}^5$ are the parameters to be identified.

\par Performance of identified models shall be quantified by the normalized simulation error, 
$\frac{\sum_t|\tilde{y}_t-y_t|^2}{\sum_t|\tilde{y}_t|^2}$, where $y_t$ denotes the simulated output of the model and $\tilde{y}_t$ denotes measured output from the system of interest.



%

\subsection{Comparison to RIE and equation error}\label{sec:msd_rie}
We first compare the Lagrangian relaxation approach to two other methods that utilize the same model structure but alternative convex surrogates for simulation error.
The first is minimization of the Local Robust Identification Error (RIE) \cite{Tobenkin2010}, 
which also gives a convex upper bound on simulation error, and was developed as a tractable approximation to Lagrangian relaxation.
The second is minimization of equation error (EE), i.e.,
\begin{equation}\label{eq:minee}
\min_\theta \ \sum_{t=1}^{T}|\eta_t|^2+\sum_{t=1}^{T-1}|\epsilon_t|^2 \text{ s.t. } E(x)+E(x)'\in\text{SOS},
\end{equation}
where the SOS constraint ensures that the identified model is well-posed (i.e. $e(\cdot)$ is a bijection). Equation error is a form of one-step-ahead prediction error, frequently used in system identification \cite{Ljung1999} and, for the case of linear systems, is exactly the algorithm of \cite{lacy2003subspace}.

\begin{figure}
\centering
\subfloat[Simulation error on training data (30 realizations).]
{\includegraphics[width=8.6cm]{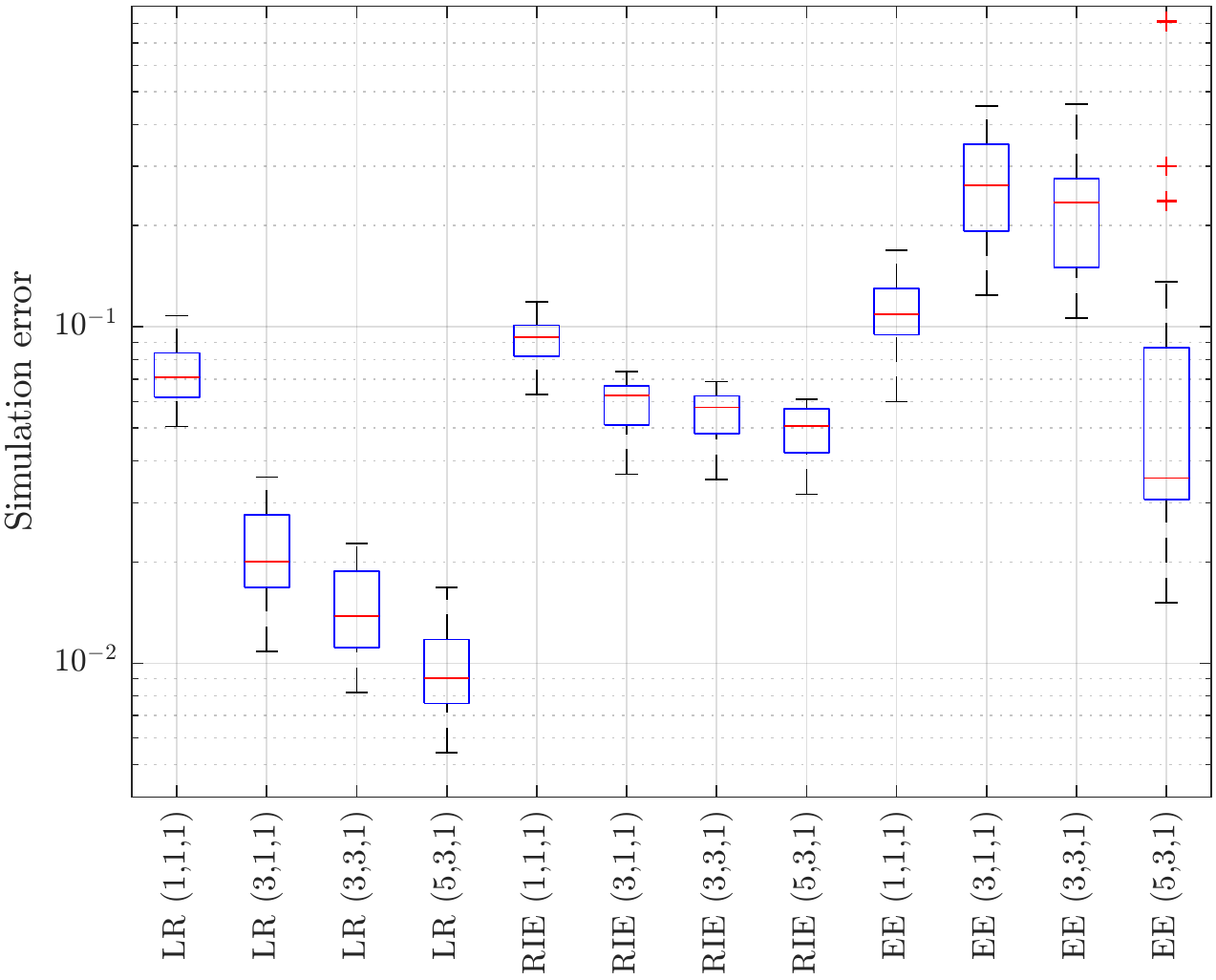}}\\
\subfloat[Simulation error on validation data.]
{\includegraphics[width=8.6cm]{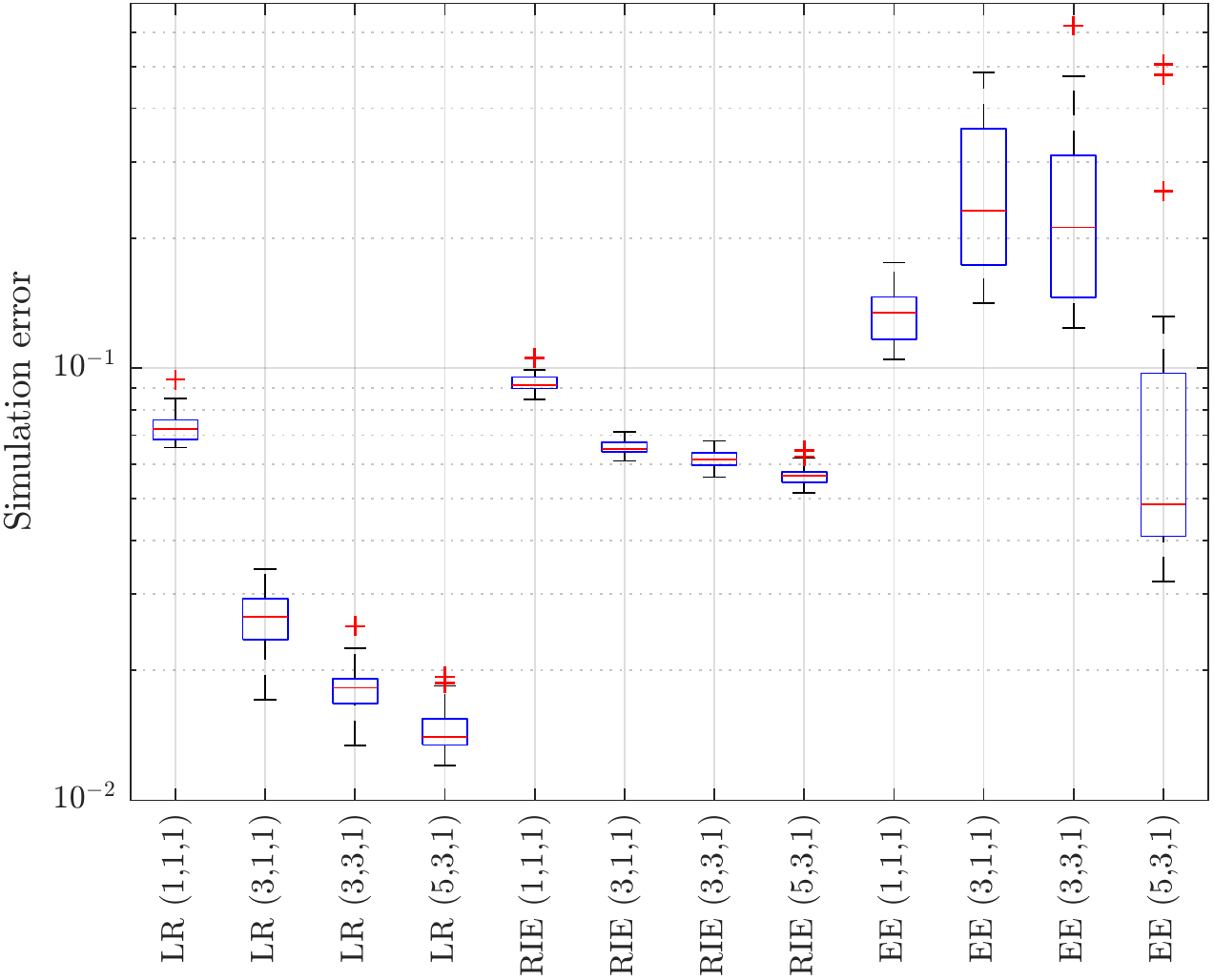}}
\caption{Comparison of different methods for fitting polynomial state-space models: Lagrangian relaxation (proposed method, LR); Local Robust Identification Error (RIE), c.f. \cite{Tobenkin2010}; equation error (EE), c.f. (\ref{eq:minee}). 
Parenthesized numbers denote the degrees of the polynomials (e,f,g) for models of the form (\ref{eq:lr_model_cs}). Refer to Section \ref{sec:msd} for experimental details.}
\label{fig:boxplot_lr_rie_ee}
\end{figure}

Identified models are of the form (\ref{eq:lr_model_cs}), with $\lbrace f_i(x,u)\rbrace_{i=1}^4$ affine in $u$.
The results are presented in Figure~\ref{fig:boxplot_lr_rie_ee}, for models of increasing complexity.
Figure~\ref{fig:boxplot_lr_rie_ee}(a) depicts performance on training data, for 30 different training data realizations.
Figure~\ref{fig:boxplot_lr_rie_ee}(b) plots the performance of these 30 different instances of each model, for a single realization of validation data. Computation times are listed in Table \ref{tab:lr_rie_ee_times}.

\par It is clear that in terms of model fidelity, LR is the best, followed by RIE, with EE worst. In terms of computation time, the ranking is reversed. This is perhaps unsurprising: minimizing EE is essentially least-squares with a small SDP constraint, and RIE was proposed as a simpler alternative to LR in \cite{tobenkin2017convex}.

We observe that for both LR and RIE, model fidelity improves (i.e. simulation error decreases) monotonically with increasing model complexity.
In contrast, performance of models fit by minimization of EE is more erratic and exhibits large variance. EE is susceptible to the well-known bias-variance tradeoff: comparing EE(1,1,1) to EE(5,3,1), we see that increasing model complexity reduces median error at the expense of large variance. In this situation, a standard remedy would be regressor pruning \cite{billings2013nonlinear}.
Both LR and RIE models achieve lower median error as model complexity increases without any increase in variance.
It should be noted that the increase in complexity from (3,3,1) to (5,3,1) is significant; these models contain 271 and 1785 parameters, respectively.
Given that only 2000 
datapoints were used for identification, 
this is evidence of the regularizing effect of model stability constraints and ``robust'' simulation error bounds.

\begin{table}
\caption{Mean computation times (in seconds, to 3 sig. fig) for the methods applied in the 30 experimental trials depicted in Figure~\ref{fig:boxplot_lr_rie_ee}. The parenthesized numbers refer to degree of $e,f,g$, respectively.}
\label{tab:lr_rie_ee_times}
\begin{center}
\bgroup
\def\arraystretch{1.0}
\begin{tabular}[c]{l|cccc}
\hline
Model & $(1,1,1)$ & $(3,1,1)$ & $(3,3,1)$ & $(5,3,1)$  \\ \hline
LR    & 4.45  & 44.4 & 67.3 & 1280 \\
RIE   & 9.47  & 20.6 & 26.4 & 172 \\
EE    & $3.60\times10^{-3}$  & $4.19\times10^{-2}$ & $4.25\times10^{-2}$ & 4.28 \\
\hline
\end{tabular}
\egroup
\end{center}
\end{table}

\subsection{Comparison to Nonlinear ARX}\label{sec:msd_nlarx}
Next, we compare our algorithm a standard approach: Nonlinear AutoRegressive models with eXogenous inputs (NARX), as implemented in the Matlab System Identification Toolbox. In particular, we compare the following identification methods:
\begin{itemize}
\item \emph{LR} - The proposed Lagrangian relaxation algorithm, applied to a model of the form (\ref{eq:lr_model_cs}) with (e,f,g) of degree (3,3,1) respectively. 
\item \emph{Poly} - 
Nonlinear regressors are all monomials in  $\lbrace\tilde{s}_n^{(1)},\tilde{s}_n^{(2)}\rbrace$ (for $n=t-1,t-2$) up to degree 3; 
\cfont{focus} = \cfont{simulation}.
\item \sigss\ - sigmoid nonlinearity; \cfont{nlreg}=\cfont{search} to select regressors; \cfont{focus} = \cfont{prediction}. 
\item \sig\ - sigmoid nonlinearity; all nonlinear regressors used; \cfont{focus} = \cfont{simulation}.
\item \wavess\ - wavelet nonlinearity; \cfont{nlreg}=\cfont{search} to select regressors; \cfont{focus} = \cfont{prediction}. 
\item \wave\ - wavelet nonlinearity; all nonlinear regressors used; \cfont{focus} = \cfont{simulation}. 
\end{itemize}

Each NARX model uses six regressors 
$\lbrace y_n^{(1)},y_n^{(2)},u_n\rbrace_{n=t-1}^{t-2}$,
with $\lbrace y_n^{(i)},u_n\rbrace=\lbrace \tilde{s}_n^{(i)},\tilde{u}_n\rbrace$ for training. 
The \cfont{focus} property was set so as to produce the best performance for each model.
This was important for the \emph{Poly} model, where \cfont{simulation} performed much better than \cfont{prediction}, but less so for the others, where the \cfont{focus} property had little influence. 

\begin{figure}
\centering
\subfloat[Training data, 30 models per method.]
{\includegraphics[width=8.6cm]{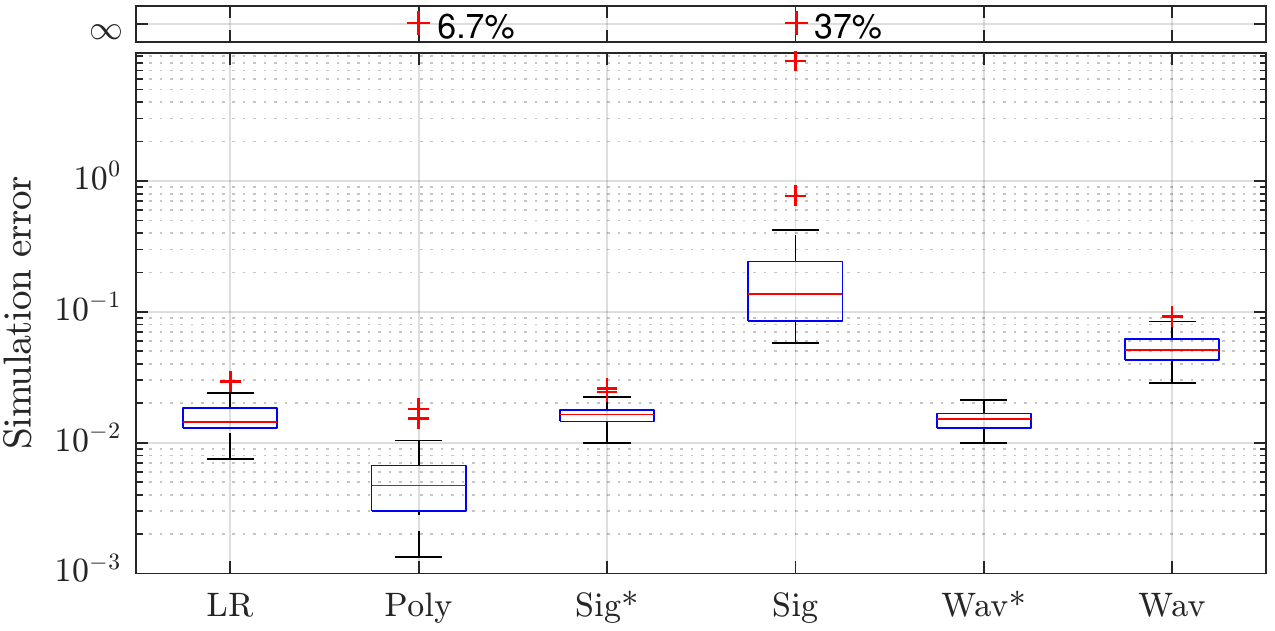}}\\
\subfloat[Validation data, 30 models per method.]
{\includegraphics[width=8.6cm]{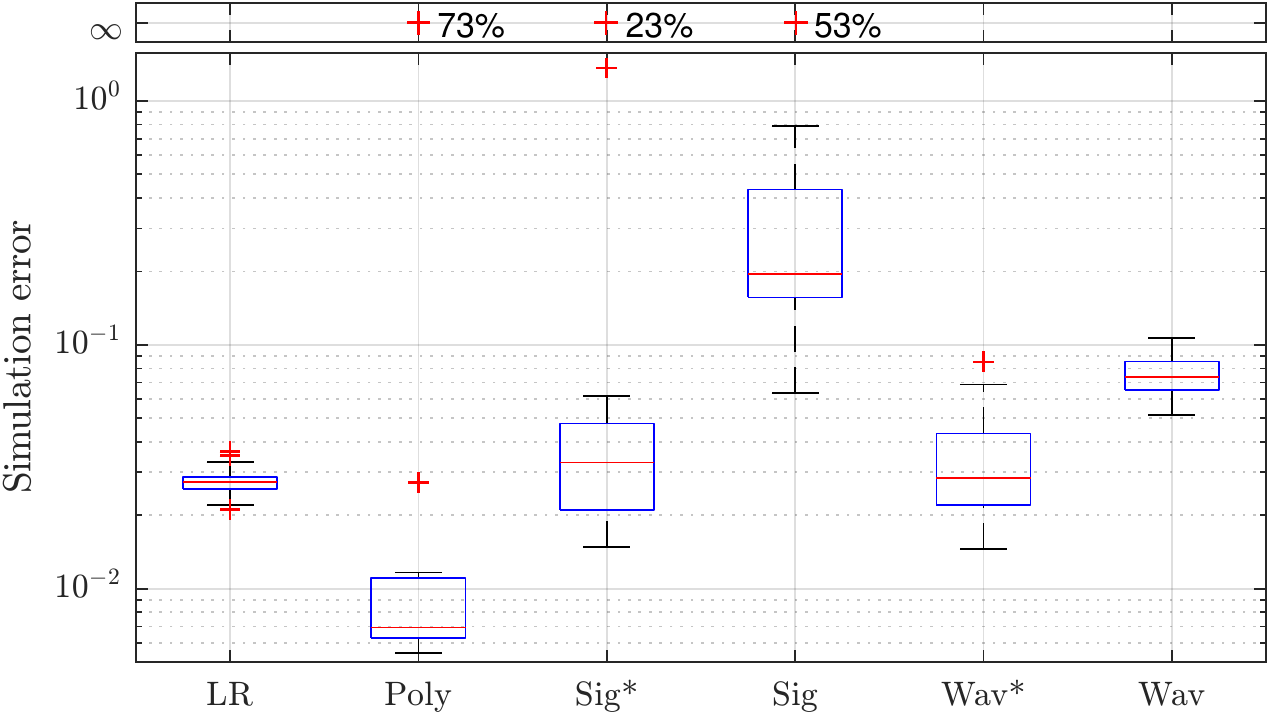}}\\
\caption{Comparison of proposed method (LR) to various nonlinear ARX models; c.f. Section \ref{sec:msd_nlarx} for a complete description of the models and methods. 
The percentages at infinite error denote the proportion of trials for which the simulated model diverged.}
\label{fig:boxplot_lr_narx}
\end{figure}

For each of the methods tested, 30 models were attained by fitting to 30 randomly generated training datasets; c.f. \mysection \ref{sec:msd}. 
Performance of these models on training data is depicted in Figure~\ref{fig:boxplot_lr_narx}(a). 
For validation, we randomly generate a \emph{single} new dataset and compute the simulation error of each of the 30 models; the results are presented in Figure~\ref{fig:boxplot_lr_narx}(b).
Many NARX models were unstable, and the simulations diverged. To keep the scale of Figure~\ref{fig:boxplot_lr_narx} meaningful, we collect these at the top as $\infty$ simulation error, and the box plots are generated using only the stable models. Note that the {\em same models} are being simulated in Figures \ref{fig:boxplot_lr_narx} (a) and (b), but different proportions of models were divergent. This is because (local) stability of a nonlinear model is trajectory-dependent. On the other hand, the global incremental stability constraint \eqref{eq:stabilityCond} for LR ensures stability for {\em all} possible inputs.


\par Some interesting observations can be made from Figure~\ref{fig:boxplot_lr_narx}.
Foremost, we note that \emph{LR} outperforms NARX, achieving the lower median error than all other methods. The apparent lower median of \emph{Poly} is is not a real effect: since 73\% of models diverged it could be said that for  \emph{Poly} the median simulation error is infinity.

The computation times are recorded in Table \ref{tab:narxtimes}. 
Computationally, \emph{LR} is comparable with \sig\ and \wave, although \emph{LR} achieves significantly lower (i.e. better) simulation error.
Only \wavess \ has similar median error to  \emph{LR}, though with larger variance, but it took around 30 times longer to compute due to the costly subset selection process.
Before moving on, we note that even better performance can be attained by LR, at the expense of greater computational effort, if we are willing to use a more complicated model, e.g. LR(5,3,1) in Figure~\ref{fig:boxplot_lr_rie_ee}. Notice that LR(5,3,1) is still twice as fast to fit compared to \wavess, c.f. Table~\ref{tab:lr_rie_ee_times}. 


\begin{table}
\caption{Computation times (in seconds, to 3 sig. fig.) for the methods applied in the 30 experimental trials depicted in Figure~\ref{fig:boxplot_lr_narx}.}
\label{tab:narxtimes}
\begin{center}
\bgroup
\def\arraystretch{1.5}
\begin{tabular}[c]{l|cccccc}
\hline
Method 		& \emph{LR}   & \emph{Poly} & \sigss    & \sig 	    & \wavess & \wave  \\ \hline
Mean 		& 63.2        & 1690        & 4930 		& 58.3 		& 2080    & 53.7 \\
Std. Dev. 	& 3.87        & 486         & 64.9 		& 28.1  	& 12.8    & 43.1 \\
\hline
\end{tabular}
\egroup
\end{center}
\end{table}

Comparing \emph{LR} to the subset selection methods \emph{Sigmoid$^*$} and \emph{Wavelet$^*$}, we observe that the \emph{variance} of simulation error on validation data is much lower for \emph{LR}.
We suggest that this is due to the large variation in the structure (i.e. selected regressors) of models from subset selection.
 Table \ref{tab:regressors} reports the frequency with which individual regressors were chosen by Matlab's subset selection algorithm. 
Notice that there isn't a single regressor that was selected in 100\% of trials. Since subset selection is inherently nonsmooth, and small variations in the training data can lead to large differences in model structure (i.e. selected regressors), having an adverse effect on the ability of these models to generalize. 
By contrast, our proposed LR algorithm involves a minimizing a {\em smooth convex} function over a convex set, and small changes in the problem data do not result in large changes in the identified model.


\begin{table}
\caption{Frequency with which certain nonlinear regressors were chosen by Matlab's subset selection algorithm (i.e. \texttt{nlreg} set to \texttt{search}) during the 30 experimental trials depicted in Figure~\ref{fig:boxplot_lr_narx}.}
\label{tab:regressors}
\begin{center}
\bgroup
\def\arraystretch{1.5}
\begin{tabular}[c]{l|cccccc}
\hline
Regressor & $y^{(1)}_{t-1}$ & $y^{(1)}_{t-2}$ & $y^{(2)}_{t-1}$ & $y^{(2)}_{t-2}$ & $u_{t-1}$ & $u_{t-2}$ \\ \hline
Wavelet$^*$, $y_1$ & 77\%  & 0\% & 17\% & 3\% & 7\% & 20\% \\
Wavelet$^*$, $y_2$ & 37\% & 27\% & 30\% & 13\% & 93\% & 97\% \\\hline
Sigmoid$^*$, $y_1$ & 80\%  & 10\% & 17\% & 10\% & 30\% & 33\% \\
Sigmoid$^*$, $y_2$ & 33\%  & 43\% & 27\% & 30\% & 67\% & 57\% \\
\hline
\end{tabular}
\egroup
\end{center}
\end{table}

\begin{figure}
\centering
\includegraphics[width=8cm]{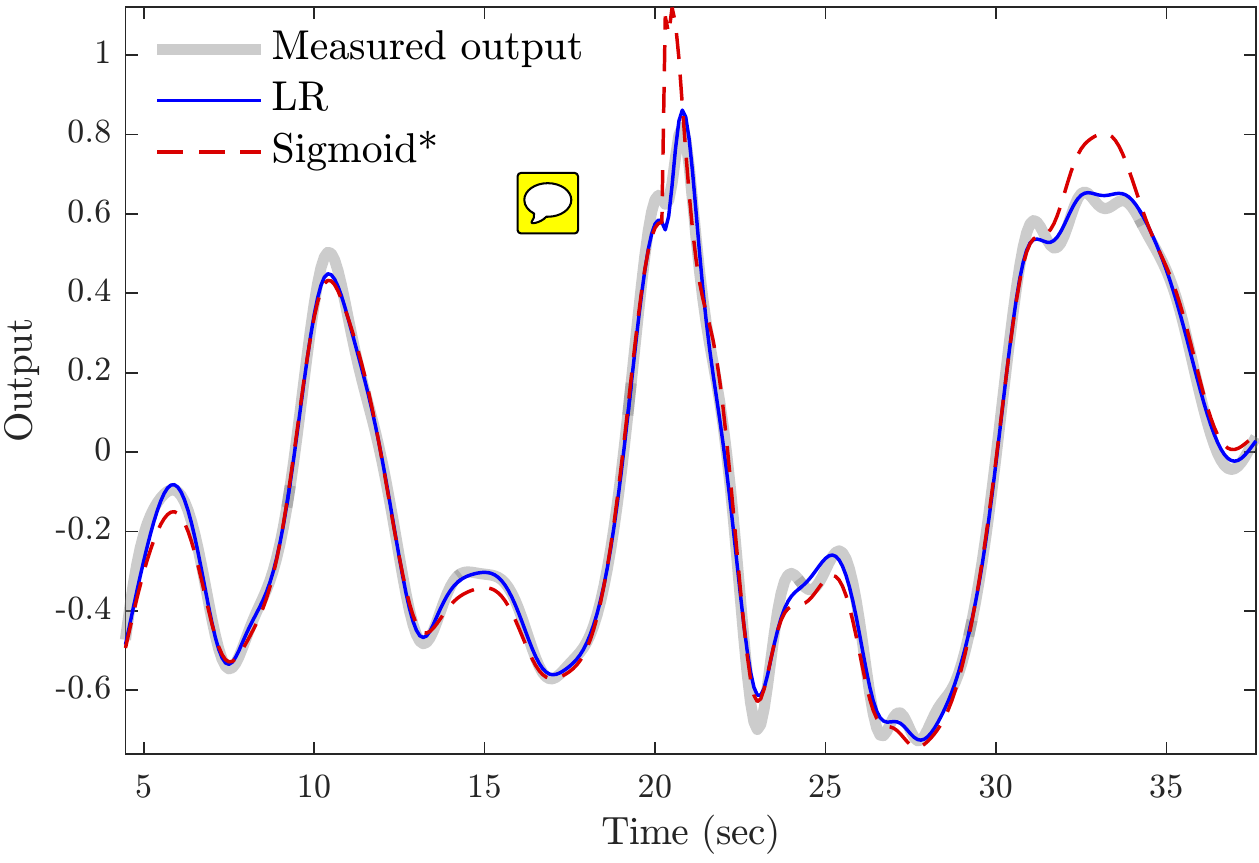}
\caption{Simulated performance on validation data for one of the trials in Figure \ref{fig:boxplot_lr_narx}.
\emph{LR} denotes a $4\myth$ order state-space model fit with our proposed algorithm.
\emph{Sigmoid$^*$} denotes a nonlinear ARX model with sigmoid net nonlinearity and regressors chosen automatically by Matlab's subset selection algorithm; see \mysection \ref{sec:msd_nlarx} for details. 
Normalized simulation error for \emph{LR} and \emph{Sigmoid$^*$} are $2.02\times10^{-2}$ and $3.71\times10^{-2}$, respectively. }
\label{fig:lrvsnarxs}
\end{figure}


\section{Case study: Two tank system}\label{sec:twotank}

In this section, we seek to model a system consisting of two interconnected tanks. 
The input is the voltage $\tilde{u}$ (V) applied to a pump, which delivers fluid to Tank One. 
Fluid then flows from an outlet in the bottom of Tank One to Tank Two. 
System output is the depth $\tilde{y}$ (m) of fluid in Tank Two. 
These signals are sampled at 5Hz to produce the discrete-time training dataset $\lbrace\tilde{u}_t,\tilde{y}_t\rbrace_{t=1}^T$, where $T=10^3$.
For further details and access to the problem data, c.f. \cite{twotank}.

\par We compare our proposed Lagrangian relaxation method to the best performing NARX model from \cite{twotank}, comprising 8 linear regressors $\lbrace\tilde{y}_{t-1},\dots,\tilde{y}_{t-5},\tilde{u}_{t-1},\dots,\tilde{u}_{t-3}\rbrace$ and 2 nonlinear regressors $\lbrace\tilde{y}_{t-4},\tilde{u}_{t-3}\rbrace$ with 12 unit wavelet nonlinearities. The polynomial model fit with Lagrangian relaxation is of the form (\ref{eq:lr_model_cs}) with $n_x=3$, $\lbrace e_i\rbrace_{i=1}^3$ degree 5, 
$\lbrace f_i=f^x_i(x)+f_i^u(u)\rbrace_{i=1}^3$ with $\lbrace f_i^x\rbrace_{i=1}^3$ degree 3 and $\lbrace f_i^u\rbrace_{i=1}^3$ degree 4, 
$g=g^x(x)+g^u(u)$ with $g^x$ degree 3 and $g^u$ degree 4. 
To estimate the internal states $\lbrace\tilde{x}_t\rbrace_{t=1}^T$, used in the construction of the Lagrange multipliers, we apply the subspace algorithm of \cite[Section 4.3.1]{van2012subspace}, with $n_x=3$.

\par
\begin{figure}
\centering
\subfloat[Simulated output on training data.]{\includegraphics[width=7cm]{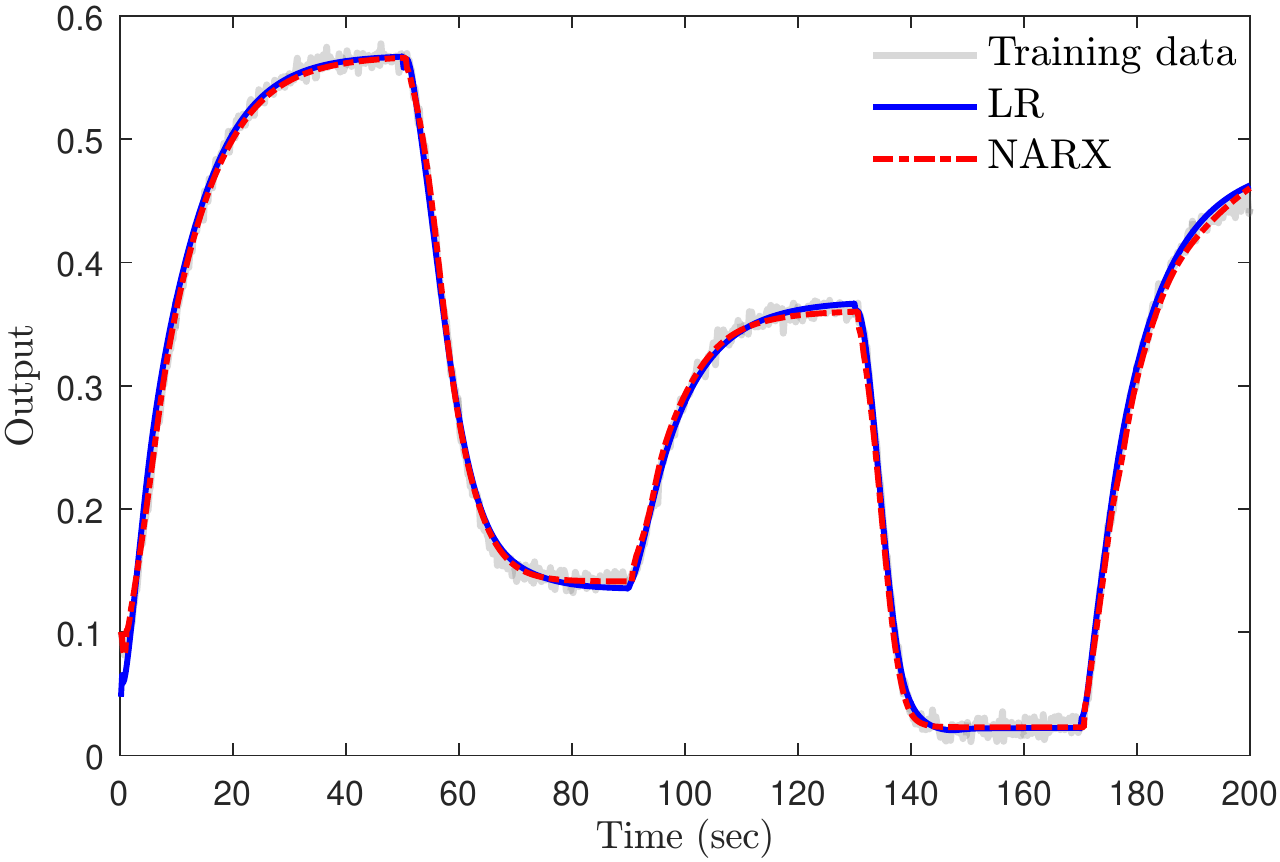}}\\
\subfloat[Simulated output on validation data.]{\includegraphics[width=7cm]{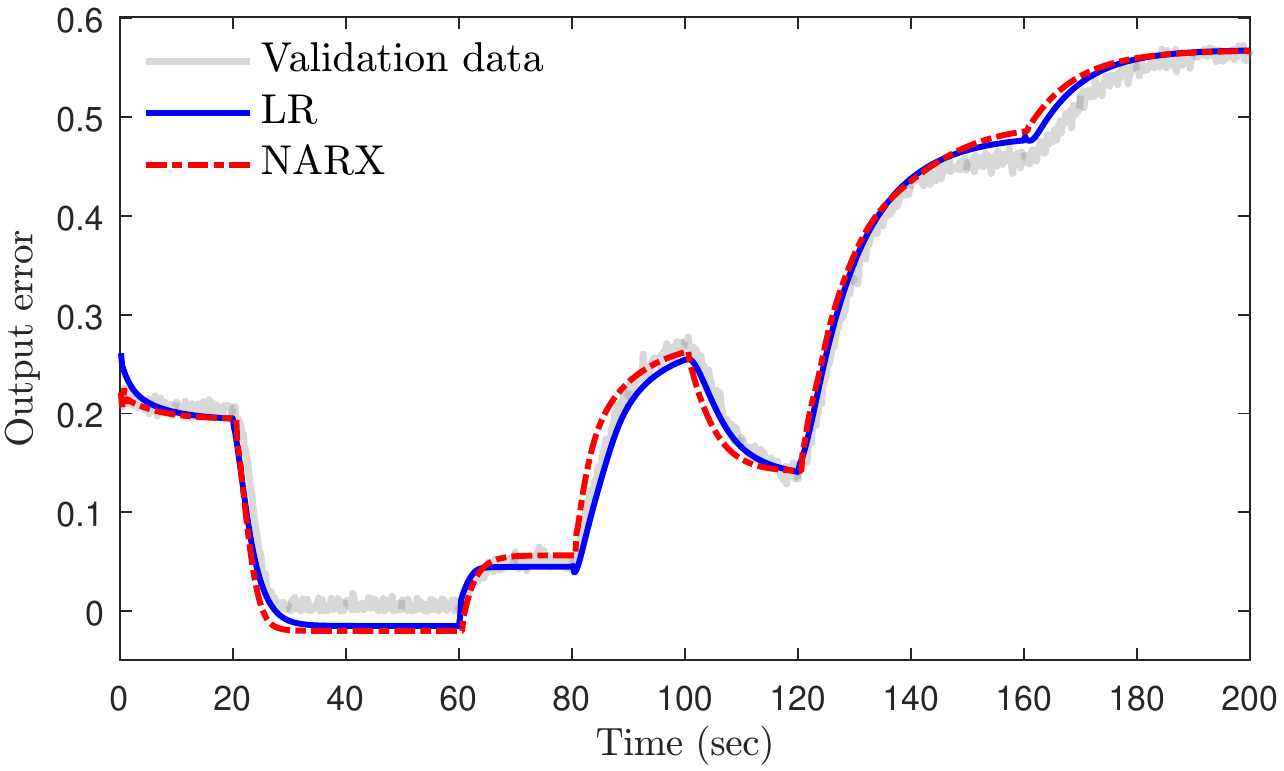}}
\caption{Simulated performance for a $3^{\textup{rd}}$ order state-space model fit with our proposed algorithm compared to a NARX model; see Section \ref{sec:twotank} for details. True data is collected from a two tank system \cite{twotank}.}
\label{fig:twotank}
\end{figure}

 The simulated performance of each model is depicted in Figure \ref{fig:twotank} and recorded in Table \ref{tab:twotank}. We observe that LR performs significantly better (49\% improvement) on validation data, compared to the best NARX model.  

\begin{table}
\caption{Normalized simulation error for training and validation data from the two tank system.}
\label{tab:twotank}
\begin{center}
\bgroup
\def\arraystretch{1.25}
\begin{tabular}[c]{l|cc}
\hline
Method      & LR                  & NARX   \\\hline 
Training    & $3.21\times10^{-4}$ & $3.62\times10^{-4}$    \\
Validation  & $2.52\times10^{-3}$ & $4.91\times10^{-3}$   \\
\hline
\end{tabular}
\egroup
\end{center}
\end{table}

%
%

\section{Case study: bias in linear  identification}\label{sec:subspace}

\par
\begin{figure}
\begin{center}
\includegraphics[width=8.4cm]{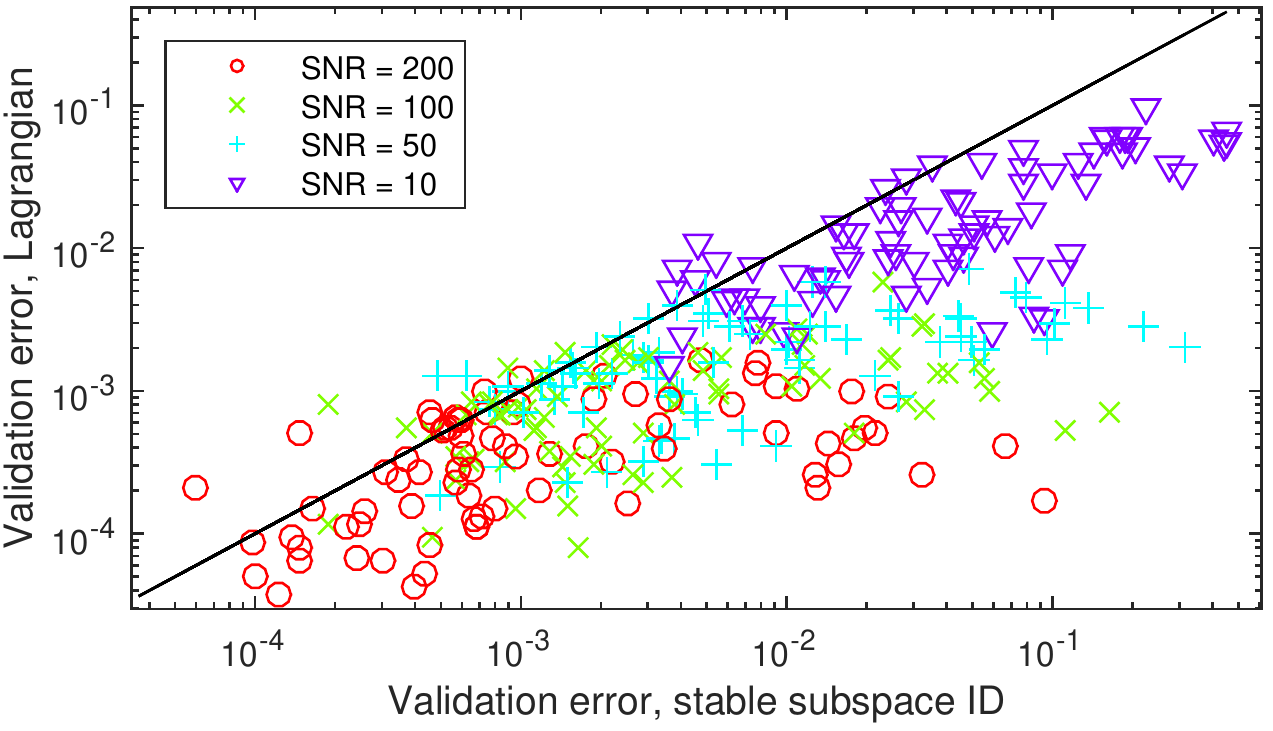}    
\caption{Performance of \ours \ compared with stable subspace ID for the identification of forty $8\myth$ order SISO models, randomly generated by Matlab's \texttt{drss} function. }
\label{fig:randomModels_1}
\end{center}
\end{figure}

 To illustrate the performance of \ours \ on a wide variety of linear models, we first conducted the following numerical experiment: Matlab's \texttt{drss} function was used to randomly generate forty $8\myth$ order LTI SISO systems. 
Each system was excited with white noise and simulated for $T=400$ time steps to generate input/output data $\lbrace \tilde{u}_t,\tilde{y}_t\rbrace_{t=1}^T$. 
The algorithm of \cite{moonen1993subspace} was used obtain an approximate state sequence $\lbrace\tilde{x}_t\rbrace_{t=1}^T$ in a balanced basis. 
We then fit $8\myth$ order linear models to the data using two methods: i) \ours, and 
ii) minimization of equation error, weighted by $P\in\sym_{++}$, subject to model stability constraints, i.e.,
\begin{align*}
\min_\theta& \ \sum_{t=1}^{T}|\ty_t-C\tx_t-D\tu_t|^2  + |P\tx_{t+1}-\xapa\tx_t-\xapb\tu_t|^2, \\
\textup{s.t.}& \quad \left[\begin{array}{cc}
P - \postol I & \xapa \\ \xapa' & P
\end{array}\right]\mgeq0,
\end{align*}
where $\theta=\lbrace P,\xapa,\xapb,C,D\rbrace$ and $\postol>0$. A stable LTI model can then be recovered as $A=P^{-1}\xapa$ and $B=P^{-1}\xapb$.
This method is henceforth referred to as `stable subspace ID'. 
This process was repeated eight times for each model, over four different SNRs. 
The results of this experiment are shown in Figure \ref{fig:randomModels_1}, which records the validation error of each identified model. 
It is clear that models identified with \ours \ outperform those from stable subspace ID in the majority (86\%) of trials.  

\begin{figure}
\centering
\subfloat[$8\myth$ order model fit to $8\myth$ order true system.]{
\includegraphics[width=8.4cm]{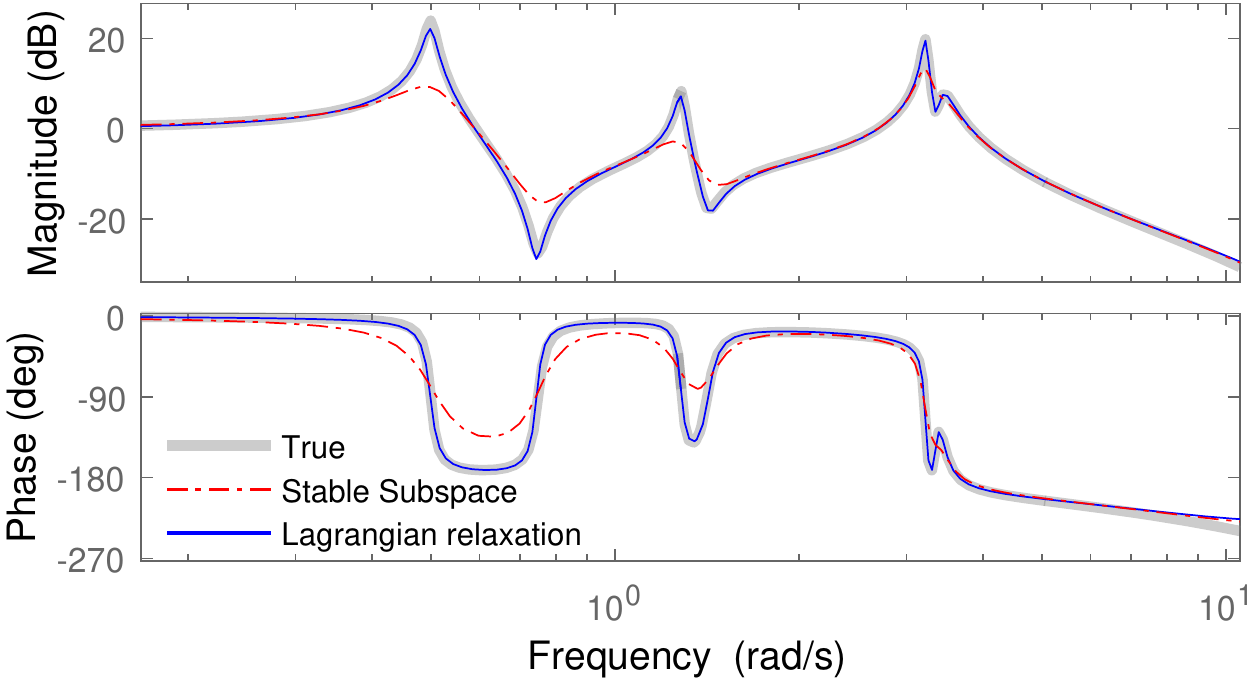}} \\   
\subfloat[$8\myth$ order model fit to $12\myth$ order true system (undermodeling).]{
\includegraphics[width=8.4cm]{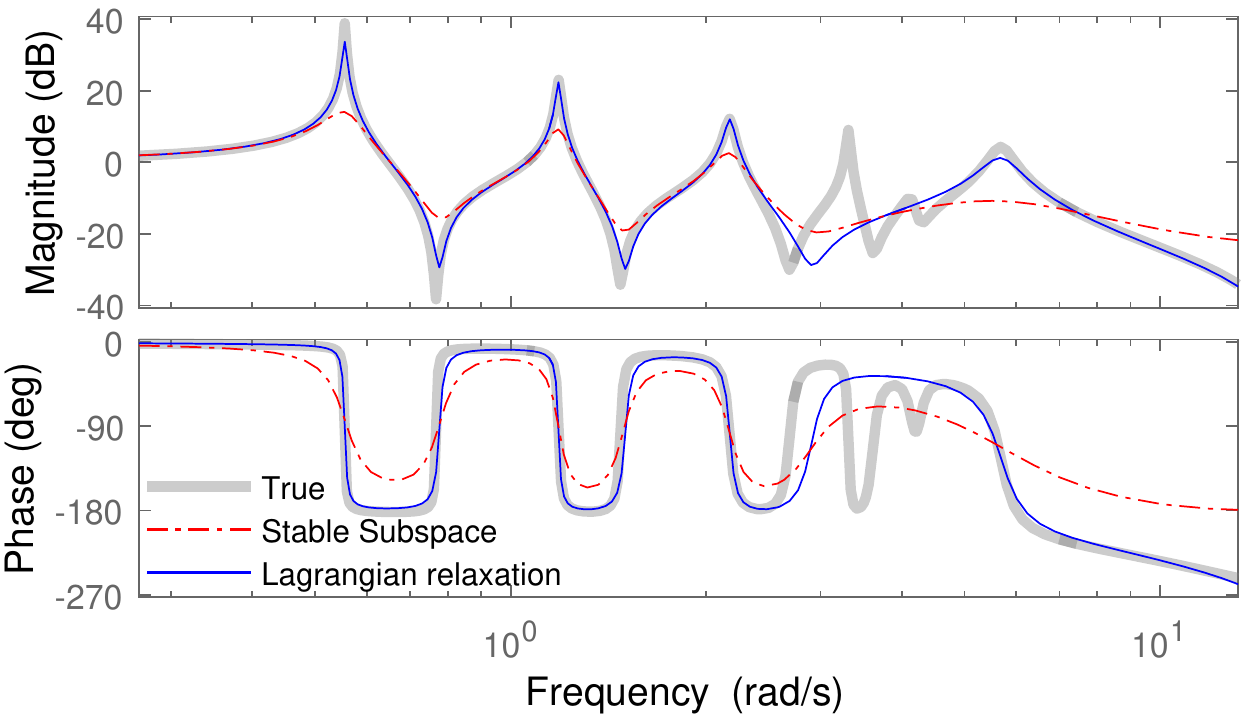}} 
\caption{Bode plots for the true flexible beam model (gray), and $8\myth$ order models identified by Lagrangian relaxation (blue) and stable subspace ID (red). In (a), the true system is $8\myth$ order, while in (b) the true system is $12\myth$ order; i.e. undermodeling is present. The output SNR was 100 (20dB).}
\label{fig:bode}
\end{figure}

\begin{figure}
\centering
\subfloat[Stable subspace ID.]{\includegraphics[width=4cm]{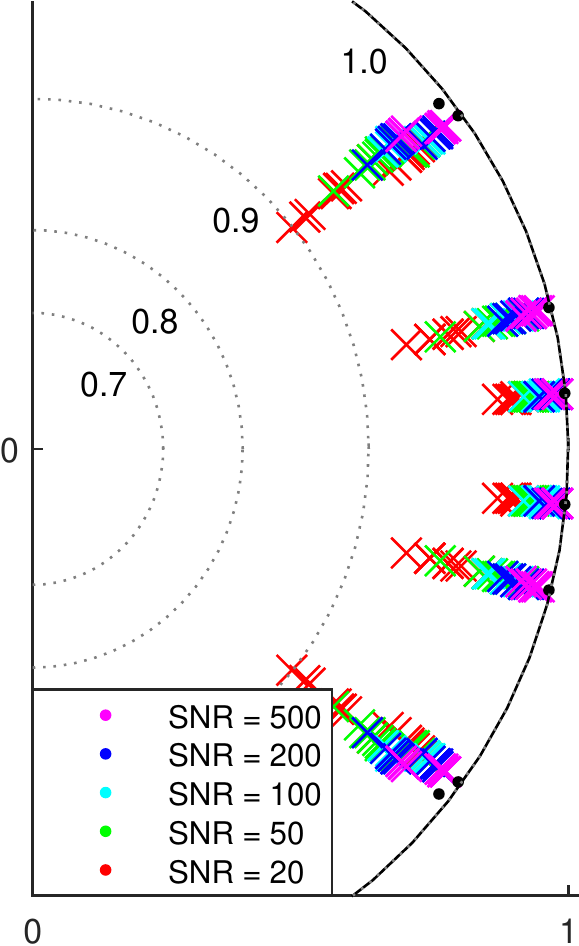}}
\subfloat[Lagrangian relaxation.]{\includegraphics[width=4cm]{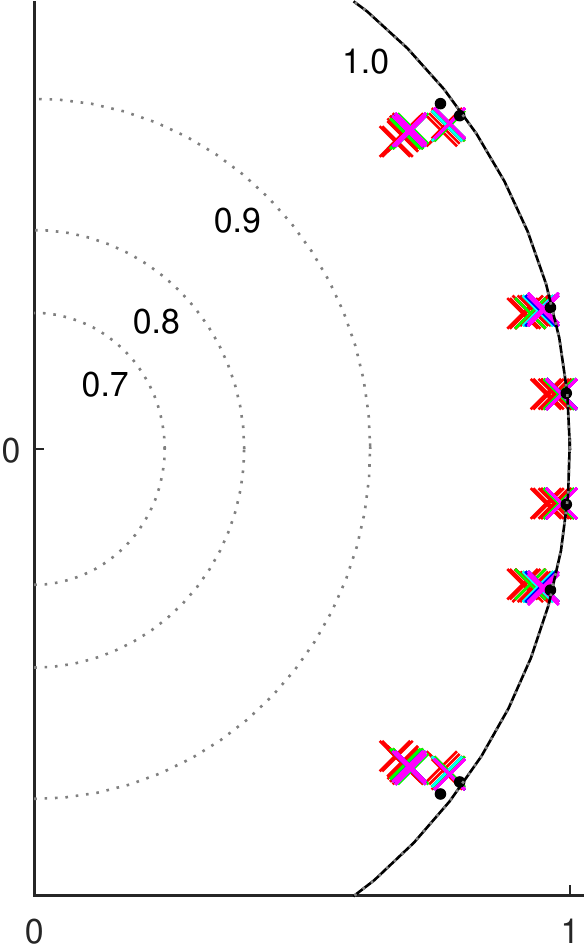}}
\caption{Pole locations of $8\myth$ order models fit to an $8\myth$ order flexible beam; c.f. Figure~\ref{fig:bode} for the Bode plot. The small dots denote the poles of the true model, `$\times$' the poles of identified models.}
\label{fig:poles_1}
\end{figure}


It has been observed by several authors that guaranteeing stability  in system identification is often associated with a bias towards models that are ``too stable'' \cite{maciejowski1995guaranteed,Manchester2012,tobenkin2014} 
and \cite{lacy2003subspace}. To gain further insight into this effect, we consider identification of a flexible beam, which serves as a useful model of cantilever structures arising in many engineering applications. 
In particular, we fit $8\myth$ order models to a $8\myth$ order (4-link) beam; the Bode plot for this system is given in Figure~\ref{fig:bode}(a). 
The subspace algorithm \cite{moonen1993subspace} was used obtain an approximate state sequence $\lbrace\tilde{x}_t\rbrace_{t=1}^T$. 
Figure~\ref{fig:poles_1} plots identified pole locations for decreasing SNR.  
Observe that the poles of models identified by stable subspace ID \cite{lacy2003subspace} have been shifted considerably towards the center of the unit circle, compared to those of the models from Lagrangian relaxation.

Figure~\ref{fig:bode}(a) presents Bode plots for identified models from one of these trials. 
The inability of the model from stable subspace ID to capture the resonant peaks -- and associated phase shifts -- is a consequence of the poles being pulled in towards the origin.
 
In most real applications, there is some degree of {\em undermodeling}: i.e. the model identified is of lower order than the true system. To examine performance in this situation,  we repeated the above experiments but fit $8\myth$ order models is fit to data from a $12\myth$ order system, representing a six-link beam. The resulting Bode plots are shown in Figure~\ref{fig:bode}(b). It is clear that LR does a good job of capturing four resonant peaks (as expected with an $8\myth$ order model), and a reasonable job of interpolating through the remaining two. The particular peaks that are captured depend on the spectra of the forcing input. Stable Subspace again fails to capture the resonance.

\section{Conclusion}\label{sec:conc}
We have developed an interior point algorithm for nonlinear system identification with guaranteed stability via Lagrangian relaxation, which takes advantage of special structure to reduce the computational complexity 
(of each Newton step) 
from cubic to linear in the data length compared to a generic SDP solver.
A straightforward Matlab implementation of this algorithm was shown empirically to achieve the same reduction in computation time (i.e., cubic to linear) compared to a highly optimized commercial solver. 
Equipped with this specialized algorithm, we demonstrate that models fit by the proposed method generalize to new datasets better than: a) models of the same structure fit by least squares without stability constraints, and b) nonlinear ARX models. 
We interpret this as evidence for the apparent regularizing effect of stability constraints and robust fidelity bounds.

%
%
%
%
%

\ifCLASSOPTIONcaptionsoff
  \newpage
\fi

\bibliographystyle{IEEEtran}
\bibliography{lrj_bib}

\vspace{-1em}

\begin{IEEEbiographynophoto}{Jack Umenberger}
received the B.E. (Hons 1) degree in Mechatronics Engineering from The University of Sydney, Australia, in 2012. He is currently a PhD candidate at The University of Sydney. His research interests include data driven modeling of dynamical systems, and motion planning and control in robotics applications.
\end{IEEEbiographynophoto}

\vspace{-1em}

\begin{IEEEbiographynophoto}{Ian R. Manchester}
received the B.E. (Hons 1) and Ph.D. degrees in Electrical Engineering from the University of New South Wales, Australia, in 2002 and 2006, respectively. He has held research positions at Umea University, Sweden, and Massachusetts Institute of Technology, USA. In 2012 he joined the faculty at the University of Sydney, where he is currently Associate Professor of Mechatronic Engineering, and a member of the Australian Centre for Field Robotics (ACFR). His current research interests include optimization methods for nonlinear system analysis, identification, and control, and applications in robotics and biomedical engineering. 
\end{IEEEbiographynophoto}

\vfill




\end{document}